\documentclass[aps,pra,10pt,twocolumn,notitlepage]{revtex4-2}
\usepackage{graphicx}
\usepackage{amssymb}
\usepackage{amsfonts}
\usepackage{amsmath}
\usepackage{amsbsy}
\usepackage{natbib}
\usepackage{comment}
\usepackage[colorlinks=true, urlcolor=blue, citecolor=blue,
linkcolor=blue]{hyperref}
\usepackage{color}
\usepackage[utf8]{inputenc}
\usepackage[T1]{fontenc}
\usepackage{amsthm}
\usepackage{mathtools}
\usepackage{scalerel}

\def\doubleunderline#1{\underline{\underline{#1}}}
\usepackage{lmodern}
\usepackage{graphicx}
\usepackage{mathtools}

\begin{document}

\title{Electrons   in intense  laser with local  phase, polarization, and skyrmionic textures}

\author{Jonas W\"{a}tzel and Jamal Berakdar}
\affiliation{Institute for Physics, Martin-Luther-University Halle-Wittenberg, 06099 Halle, Germany}
%

\begin{abstract}
Laser fields can be shaped on  sub-wavelength scale as to have a specific  distribution in  spin angular momentum,  orbital structure or topology. We study how these various features affect the strongly non-linear electron dynamics. Specifically, we derive closed expressions for the wavefunction of an unbound electron subject to a generally structured, intense laser field and  demonstrate its use for  imprinting  the orbital angular momentum of a propagating optical vortex onto photoelectrons emitted from atoms and traveling through the optical vortex. It is also shown that photoelectrons can be accelerated or momentum textured  when moving through a focused, intense laser field whose spin angular momentum is modulated as to have a radial polarization which also implies the presence of a strong electrical longitudinal component. Further results are presented on the sub-wavelength spatio-temporal imaging of a laser field topology, as demonstrated explicitly for the field's spin and orbital distributions of lossless propagating optical skyrmions sampled by photoelectrons.
\end{abstract}

\date{\today}

\maketitle

\section{Introduction}
The precise shaping of the time structure of laser pulses has been the basis for time-resolved spectroscopy and ultrafast science \cite{Zewail1645,krausz2009attosecond}. Engineering the local polarization state and/or the spatial phase of the wavefront offers further opportunities for  applications. Such local engineering of optical fields can be brought about by various means: Appropriately designed photonic elements such as waveguides \cite{born2013principles, saleh2019fundamentals} host eigenmodes with desirable polarization distribution. Also, specifically decorated plasmonic structures allow  tuning the spatial distribution of the spin (polarization) and orbital (phase) structure  of the electromagnetic field \cite{tsesses2018optical, Spektor1187, PhysRevX.9.021031, dai2019ultrafast, davis2020ultrafast, D0NR00618A, omatsu2017focus, Kerber_2018, ayuso2019synthetic}. In this way, optical skyrmions and plasmonic waves carrying orbital angular momentum were realized. While plasmonic fields overcome the limitation on the spatial resolution of diffraction-limited propagating (laser) wave in free space, the latter offers a large flexibility in tuning the frequency, intensity, and carrier-envelope phase at very low power losses. Therefore, much efforts were devoted to the spatial structuring of the polarization and wavefront phase of freely propagating waves. For instance, laser pulses with azimuthal and radial polarization \cite{born2013principles, saleh2019fundamentals} were realized in a wide frequency range \cite{Mitchell:s,zhan2009cylindrical,Erd_lyi_2008,kozawa2005generation,hernandez2017extreme}. \\
The topic in general is attracting much research recently due to the  great potential for fundamental and applied  science \cite{Rubinsztein_Dunlop_2016}. For instance, orbital (OAM) and spin (SAM) angular momentum carrying pulses can generate unidirectional charge currents \cite{Quinteiro:09, watzel2016optical, sederberg2020Vectorized} which is interesting for  opto (spin)electronic applications  \cite{PhysRevB.90.115401,Watzel:12, solyanik2019spin,PhysRevB.93.045205,KOC2015599,PhysRevB.100.115308,doi:10.10631.5027667, ji2020photocurrent}. For molecules \cite{PhysRevLett.89.143601,PhysRevA.71.055401,PhysRevLett.96.243001}, structured laser pulses are expected to yield new information, particularly on chiral and helical molecular aggregates \cite{PhysRevA.99.023837,Wozniak:19,ayuso2019synthetic}. \\ 	
For atoms, the  electron wave function is extremely localized with respect to variation in the spin or orbital parts of optical fields. Thus, at first glance, it seems that the local structure of the laser field is marginal when considering  the response of a random distribution of non-interacting atoms in the laser spot \cite{kaneyasu2017limitations}. On the other hand, the local spatial variation of the  SAM and phase of the laser-field are not diffraction-limited and may change on  the sub-wavelength scale.
However, already in the perturbative regime, one can identify an optimal position of the atom within the laser spot where the phase structure of the laser is important, and at the same time the transition probability is sizable \cite{PhysRevA.86.063812}. For a trapped cold atom, theory and experiment revealed much details on which types of bound-bound transitions are triggered by OAM-carrying light (for instance, \cite{schmiegelow2016transfer, afanasev2018experimental,PhysRevA.97.023422,Duan_2019,peshkov2016absorption}).
Interaction with OAM carrying pulses with resulting in photoemission has also been the subject of various theoretical studies  \cite{picon2010photoionization, seipt2016two, baghdasaryan2019dichroism,watzel2016discerning}. \\
Summarizing the status of knowledge on electrons in structured propagating fields, one may say that the direct ionization by an OAM carrying pulse is relatively well understood. Continuum-continuum (CC) transitions involving OAM exchange are less unstudied, however. The investigation of the two-photon transition matrix element corresponding to a conventional XUV field and an IR vortex illustrated  the impact of the (transferred) OAM on the CC phase and the associated time delay \cite{giri2020signatures}. The recent experimental and theoretical work  \cite{de2020photoelectric} (cf. in particular the supplemental materials of \cite{de2020photoelectric}) clearly highlights the importance of using the correct structured-light-matter interaction, including the longitudinal field component and also the role of the position of the atom in the laser spot. These two aspects (among others) are  inherent features of the interaction of matter with structured light and will be discussed at length within the framework of our developed theoretical model.\\
Interaction with SAM structured fields (vector beams) with atoms is much less studied. How a spin-orbitally coupled electronic system react to vector beams was addressed in Refs. \cite{watzel2019magnetoelectric, watzel2020nanostructures}. High harmonic generation (HHG) upon a strongly nonlinear driving of atoms with vector beams was reported in Refs.\, \cite{hernandez2017extreme,watzel2020multipolar,watzel2020topological}.
The interaction of atoms with optical skyrmions was recently formulated in \cite{watzel2020topological}, and nonlinear electron dynamics was simulated.\\
In this study we will deal with intense propagating (laser) fields having 	
a spatial variation in the orbital or spin or in both (such as in skyrmions) \cite{allen1992orbital, andrews2012angular,bliokh2015transverse,Barnett_2016,ALLEN1999291,Bouchard_2014}. One key goal is to derive a unified quasi-analytical description of non-linear electron dynamics in such structured laser beams.
Moreover, the derived expressions allow for the incorporation of a laser pulse with arbitrarily (but reasonably) SAM or/and OAM structured pulse. Utilizing the strong-field approximation \cite{keldysh1965ionization, faisal1973multiple, reiss1980effect, lewenstein1994theory}, the derived electron state in the presence of  structured light fields are used for the calculations of laser-induced electron emission in dependence on the optical OAM of the laser-assisting fields. In addition, it is demonstrated how intense and tightly focused SAM-structured vector beams \cite{zhan2009cylindrical} can be employed for linear momentum texturing of electronic wave packet.
%
Furthermore, we demonstrate that (photo)electron dynamics can sample the spatio-temporal structure of intense propagating optical skyrmions where the optical OAM and SAM are intertwined  \cite{tsesses2018optical,Spektor1187,dai2019ultrafast,watzel2020topological}.

%
%

\section{Mathematical description of structured propagating laser fields}

\label{sec:vortex_math}
\subsection{General considerations}
In the vicinity of the optical axis, Bessel \cite{volke2002orbital} and Laguerre-Gaussian modes \cite{allen1992orbital} exhibit locally similar functional dependencies \cite{watzel2020multipolar}. Cylindrical coordinates $\pmb{r}=\left\{ \rho,\varphi,z \right\}$ allow a convenient  description of several types of structured beams, including beams carrying OAM, radially and azimuthally polarized vector beams as well as propagating optical skyrmions. Generally, for these beams the key ingredient is the vortex vector field $\pmb{A}_{\rm OV}^{m^{(a)},\sigma_{\rm L}}(\pmb{r},t)$, whose mathematical expression is given explicitly below. $m^{(a)}$ is the topological charge with the superscript $a=\pm$ signaling the vortex chirality reflecting the direction of the embodied OAM. $\sigma_{\rm L}=\pm1$ indicates  the polarization state (direction of the SAM). For the behavior of the non-paraxial vector potential employed below,
the direction of the beam's OAM relative to SAM is important. We distinguish between the \emph{parallel} case, i.e., ${\rm sgn}(\sigma_{\rm L})=a$ and the \emph{antiparallel} case where ${\rm sgn}(\sigma_{\rm L})=-a$. Of a particular  interest is the region near the optical axis  $\rho\approx 0$ (on the scale of the beam waist $w_{\rm L}$) \cite{quinteiro2015formulation}, as discussed for instance in \cite{PhysRevA.71.055401}  for the case of OAM carrying laser beam.
\subsubsection{Parallel SAM and OAM, ${\rm sgn}(\sigma_{\rm L})=a$}
 The vector potential in the parallel case can be written as
\begin{equation}
\begin{split}
\pmb{A}^{m^+,+1}_{\rm OV}(\pmb{r},t)=&A_0F_m(\rho)e^{im\varphi} e^{i(q_zz-\omega_{\rm L} t)}\hat{e}_{\sigma_{\rm L}=+1} \\
& + {\rm c.c.}
\end{split}
\label{eq:parallel1}
\end{equation}
and
\begin{equation}
\begin{split}
\pmb{A}^{m^-,-1}_{\rm OV}(\pmb{r},t)=&(-1)^mA_0F_m(\rho)e^{-im\varphi} \\
&\times e^{i(q_zz-\omega_{\rm L} t)}\hat{e}_{\sigma_{\rm L}=-1} + {\rm c.c.} .
\end{split}
\label{eq:parallel2}
\end{equation}
A striking feature is the absence of a longitudinal component so that the approximate (i.e., for $\rho\approx 0$ ) vector potential is fully transverse. The laser field  propagates effectively along the $z$ axis with the wave vector $q_z$, the amplitude is set by  $A_0$ which determines the laser intensity and $\hat{e}_{\sigma_{\rm L}}=(\hat{e}_\rho + i\sigma_{\rm L}\hat{e}_\varphi)e^{i\sigma_{\rm L}\varphi}$ is the circular polarization vector. The dispersion relation is $$q_\perp^2+q_z^2=q_{\rm L}^2=\omega_{\rm L}^2/c^2,$$
and the radial distribution reads
$$F_m(\rho)=(q_r\rho)^m.$$
The transverse wave vector $q_\perp $ is related to the beam waist $w_{\rm L}$ as  $q_\perp \simeq 1/w_{\rm L}$. The parallel class vector potentials are fully transverse (within the adopted approximation) so that OAM and SAM are separable \cite{bliokh2015transverse}. The carried total angular momentum is $\hbar(m+\sigma_{\rm L})$.

\subsubsection{Anti-parallel SAM and OAM, ${\rm sgn}(\sigma_{\rm L})=-a$}
\begin{figure}
  \centering
  \includegraphics[width=0.9\columnwidth]{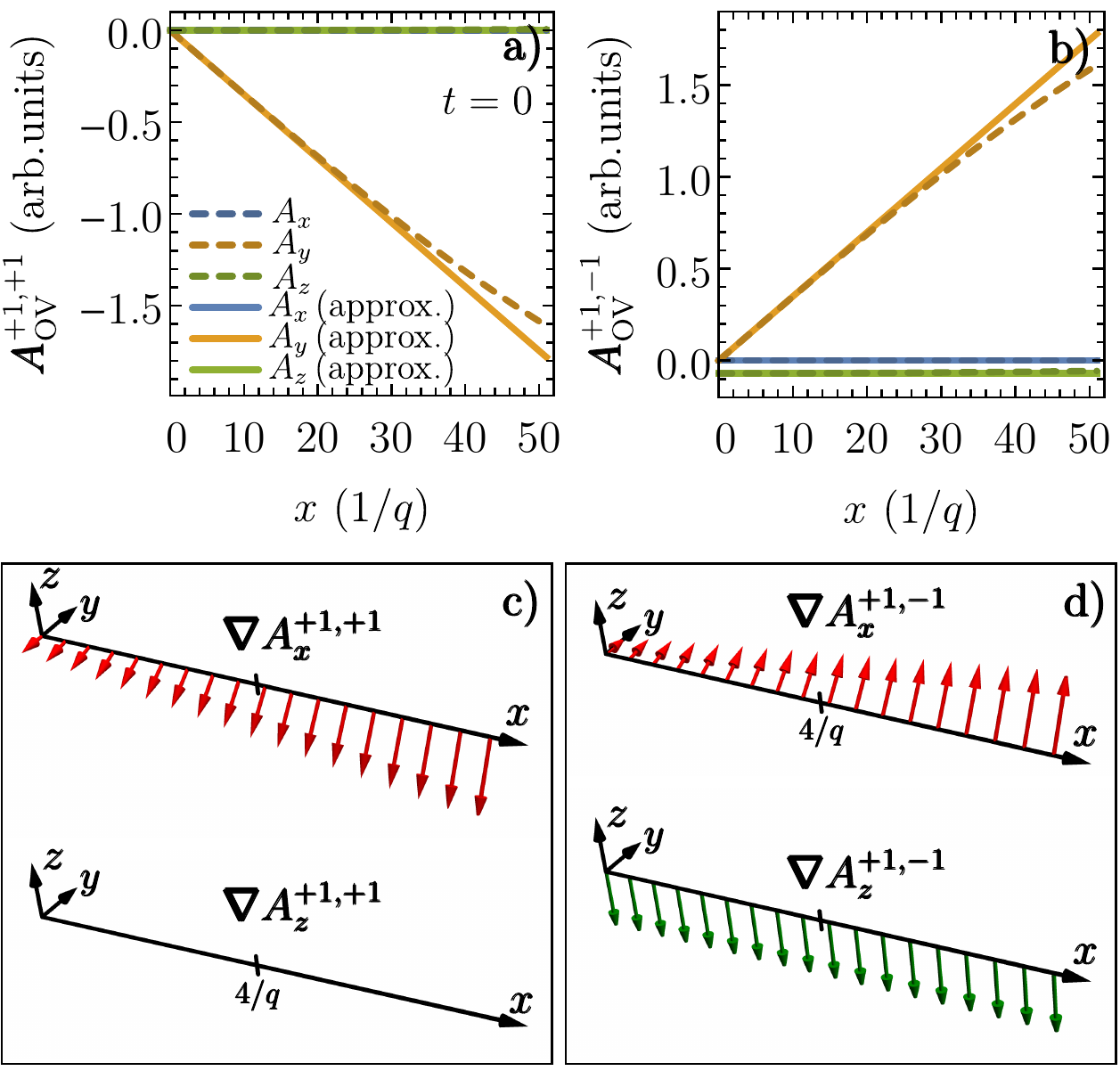}\\
  \caption{Upper row: Spatial distributions of the vector potential components corresponding to the parallel and antiparallel OAM and SAM. The vortex topological charge is $+1$ while the laser focus is set by $\alpha=\arctan(q_\perp/q_z)$ and is chosen to be $1^\circ$. Lower row: the gradients  in both cases of  the vector potentials along the $x$-axis.}
  \label{fig1}
\end{figure}

 For the antiparallel case one finds
\begin{equation}
\begin{split}
\pmb{A}^{m^+,-1}_{\rm OV}(\pmb{r},t)=&\left(F_m(\rho) \hat{e}_{\sigma_{\rm L}=-1} +i2m\frac{q_\perp}{q_z}F_{m-1}(\rho)e^{-i\varphi}\hat{e}_{z}\right) \\
& \times A_0e^{im\varphi} e^{i(q_zz-\omega_{\rm L} t)}  + {\rm c.c.}
\end{split}
\label{eq:antiparallel1}
\end{equation}
and
\begin{equation}
\begin{split}
\pmb{A}^{m^-,+1}_{\rm OV}(\pmb{r},t)=&\left(F_m(\rho)\hat{e}_{\sigma_{\rm L}=+1} + i2m\frac{q_\perp}{q_z}F_{m-1}(\rho) e^{+i\varphi}\hat{e}_{z}\right) \\ & \times (-1)^mA_0e^{-im\varphi} e^{i(q_zz-\omega_{\rm L} t)}  + {\rm c.c.}.
\end{split}
\label{eq:antiparallel2}
\end{equation}
These equations evidence the presence of a longitudinal component, {  whose  strength (relative to the transverse component) is determined by the focusing condition.}\\
For the following discussion, it is important to note that the longitudinal component scales as $r^{m-1}$ in the antiparallel case. Consequently, for $m=1$ the on-axis field does not vanish along the propagation direction. The accuracy of the chosen approximation is demonstrated in Fig.\,\ref{fig1} (first row), where the cartesian components of the vector potential function for both classes are presented. Up to distances of $25/q$, an optical vortex is well-described by the approximation given in Eqs.\,\eqref{eq:parallel1} and \eqref{eq:antiparallel1}. Note, the presence of a longitudinal component does not invalidate $\pmb{\nabla}\cdot\pmb{A}^{m^{(a)},\sigma_{\rm L}}_{\rm OV}(\pmb{r},t)=0$ for all vector potentials in Eqs.\eqref{eq:parallel1}-\eqref{eq:antiparallel2}. Due to the non-vanishing longitudinal component, OAM and SAM are not separable \cite{bliokh2015transverse}.

 \subsection{Optical vortices, polarization structured beams and propagating optical skyrmions}
{Propagating optical fields with desired  polarization and spatial phase textures can be constructed as a linear combination of the vector functions  $\pmb{A}^{m^{(a)},\sigma_{\rm L}}_{\rm OV}(\pmb{r},t)$:}
\begin{itemize}
\item A \emph{linearly polarized} beam carrying the OAM value $m^{(a)}\hbar$  can be viewed  as an optical vortex with topological charge $m^{(a)}$ and  is expressible as the superposition $\pmb{A}^{m^{(a)},+1}_{\rm OV}(\pmb{r},t)\pm\pmb{A}^{m^{(a)},-1}_{\rm OV}(\pmb{r},t)$. The vector field does not transfer a net SAM during the interaction.
\item An \emph{azimuthally polarized vector beam} \cite{zhan2009cylindrical} (AVB) can be written as the coherent sum $\pmb{A}^{+1,-1}_{\rm OV}(\pmb{r},t) + \pmb{A}^{-1,+1}_{\rm OV}(\pmb{r},t)$. The explicit expression of the vector potential is given by \cite{watzel2019magnetoelectric}
\begin{equation}\label{eq:avb}
\pmb{A}_{\rm AVB}(\pmb{r},t) = A_0q_\perp\rho\sin(q_zz-\omega_{\rm IR} t)\hat{e}_\varphi.
\end{equation}
    Note the absence of the longitudinal component.
\item  A \emph{radially polarized vector beam} (RVB)  is expressible as  $\pmb{A}^{+1,-1}_{\rm OV}(\pmb{r},t) - \pmb{A}^{-1,+1}_{\rm OV}(\pmb{r},t)$ \cite{watzel2020multipolar}. For a tightly focused beam \cite{dorn2003sharper}, it is sometimes useful to Taylor-expand the spatial radial distributions $F_{m=0}$ to second or higher orders. For instance,  $F_0(\rho)\approx 1 - (q_\perp\rho_0)^2/4$ and $F_1(\rho)=q_\perp\rho_0 - (q_\perp\rho_0)^3/8$. The corresponding vector potential reads then
\begin{equation}
\begin{split}
\pmb{A}_{\rm RVB}(\pmb{r},t) =& A_0\left[\vphantom{\frac{q}{q}} \left(q_\perp\rho-(q_\perp\rho_0)^3/8\right) \cos(q_zz-\omega_{\rm IR} t) \hat{e}_\rho \right. \\
&\left. - 2\frac{q_\perp}{q_z}\left(1-(q_\perp\rho_0)^2/4\right)\sin(q_zz-\omega_{\rm IR} t) \hat{e}_z \right],
\end{split}
\end{equation}
where a longitudinal field is present at the optical axis, and still $\pmb{\nabla}\cdot\pmb{A}_{\rm RVB}(\pmb{r},t)=0$ is sustained. As for AVB, the radially polarized vector beam does not exhibit a net OAM nor SAM. Yet, the well-defined spatial structuring of SAM does affect electron dynamics in a unique way. For instance, the AVB can act on electrons as a gauge-invariant vector potential, leading to a transient Aharonov-Bohm-type, non-dissipative current (meaning, AVB triggers a time-dependent orbital magnetic moment even if the net SAM of the field vanishes) \cite{watzel2019magnetoelectric}.
\item  An \emph{optical propagating skyrmion} \cite{watzel2020topological} reveals a rich phase and position-dependent polarization landscape. A convenient mathematical representation is  $\alpha\pmb{A}^{m_1^+,+1}_{\rm OV}(\pmb{r},t)+\beta\pmb{A}^{m_2^{-},-1}_{\rm OV}(\pmb{r},t)$ ($\alpha,\beta$ are real numbers) resulting in the vector potential
\begin{equation} \label{eq:optskr}
\begin{split}
&\pmb{A}_{\rm OS}^{m_1^+,m_2^-}(\pmb{r},t) = A_0e^{i(q_zz-\omega_{\rm IR}t-(m_2+1)\varphi)} \\
&\quad\times\left[\left( \vphantom{e^{i(m_1+2)}} \alpha e^{i(m_1+m_2+2)\varphi}(q_\perp\rho)^{m_1} + \beta(-q_\perp\rho)^{m_2} \right)\hat{e}_\rho \right. \\
&\quad+ i\left.\left( \vphantom{e^{i(m_1+2)}} \alpha e^{i(m_1+m_2+2)\varphi}(q_\perp\rho)^{m_1} - \beta(-q_\perp\rho)^{m_2} \right)\hat{e}_\varphi\right]\\
&\quad + c.c..
\end{split}
\end{equation}
The vector potential is not transverse and the carried OAM and SAM  \cite{bliokh2015transverse} are intertwined  in a way that may be characterized by a topological quantity in analogy to the skyrmion number of a magnetic skyrmion (for details on magentic skyrmions we refer to Ref. \cite{Fert2017} and references therein). In Ref.\cite{watzel2020topological}  we discussed  a possible definition of  optical skyrmionic topological index but we should note  a key difference to magnetic skyrmions. In \cite{jia2019twisting} for example, we discussed how by increasing the radius of a magnetic disc (corresponding to changing the beam waist in the laser beam) the  magnetic ordering transforms from a vortex  to a  skyrmionic state eventually reaching a uniform magnetic ordering; an OAM carrying wave   \cite{Jia2019}  may also occur. This behavior can be described within one unified, mathematically consistent picture. In contrast to   optics, in magnetism we are dealing with static vector field  stabilized by internal interactions. For linear materials, the time-average of propagating (or plasmonic)   electromagnetic fields vanishes, regardless of their spatial or spin structure. The relevance  of  geometry or topology of optical fields is manifested by  the type of the   processes they trigger  when interacting with matter \cite{ayuso2019synthetic,hernandez2017extreme,Watzeltoroid,watzel2019magnetoelectric,watzel2020multipolar,watzel2020topological}, as  illustrated below.
\end{itemize}
\section{Light-matter interaction}
The interaction of an electron with an arbitrarily structured laser field propagating in free space with  the wave vector  $k$
and described by the vector potential $\pmb {A}_L$
follows from a minimal coupling scheme. The  Langrangian density is cast as
${\cal L}={\cal L}_{mech}+{\cal L}_{field}+  \pmb {j}\cdot\pmb {A}_L -\rho_c \Phi_L$, where ${\cal L}_{mech}$ and ${\cal L}_{field}$ are the mechanical and field parts, and $\pmb{j}$ and $\rho_c$ are the current and charge densities, respectively. $\Phi_L$ is the scalar potential. Thus, the interaction of matter with the field delivers
 in general  two contributions  to the Hamiltonian. The current-current coupling term yields
  (atomic units (a.u.) are used, unless stated otherwise)
\begin{equation}
\hat{H}_{\rm CC-int}=\frac{1}{2}\left[ -i\pmb{\nabla}\cdot\pmb{A}_L(\pmb{r},t) -2i \pmb{A}_L(\pmb{r},t)\cdot\pmb{\nabla} + \pmb{A}_L^2(\pmb{r},t)\right]
\end{equation}
Even in free space the charge density couples to the laser scalar potential $\Phi_L$, where
$\partial_t \Phi_L(\pmb{r},t)=- c^2 \pmb{\nabla}\cdot\pmb{A}_L$.
It is possible to transform from this Lorenz gauge to an instantaneous  (Coulomb) gauge \cite{peshkov2017photoexcitation, de2020photoelectric} (which is adopted henceforth) by  introducing the vector potential  $\pmb{A}=\pmb{\nabla}(\pmb{\nabla}\cdot \pmb{A}_L)/k^2+ \pmb{A}_L$.
 In this gauge $\Phi(\pmb{r},t)$ does not appear in the light-matter interaction, however the longitudinal component of the vector potential $\pmb{A}$ can be decisive \cite{watzel2020multipolar,watzel2020topological} affecting $\rho_c$ via non-dipolar transitions.  Thus, denoting with  $\hat{\pmb{p}}_i$  the momentum operator of the $i^{th}$ electron, and with $\pmb{A}(\pmb{r},t)=\sum_j \pmb{A}_{\rm j}(\pmb{r},t)$  the total (sum) vector potential of all present fields $\pmb{A}_{\rm j}$, we may write
  in general   for the light-matter interaction
\begin{equation}
\hat{H}_{\rm int}(t)= \sum_i\pmb{A}(\pmb{r}_i,t)\cdot\hat{\pmb{p}}_i + \frac{1}{2}\pmb{A}^2(\pmb{r}_i,t).
\end{equation}
The expression is formally similar to the case of uniform fields but physically leads to the same effects as in Lorenz gauge such as the possible excitation of volume charge-density modes   \cite{watzel2019magnetoelectric}.\\
Figure\,\ref{fig1} shows the local gradient of the vector potential (for parallel and antiparallel  classes) in the vicinity of the optical axis. In contrast to the transversal components, the corresponding gradient remains finite when approaching  $\rho=0$. Consequently, the dynamics around the optical axis in the parallel case is mainly driven by this gradient. Note, $\pmb{\nabla}\pmb{A}_{\rm OV}^{m^+,+1}$ points into the $\varphi$-direction which is associated with the intrinsic
 phase structure of the vortex field and is proportional to $m/\rho$. In the antiparallel case, the near-axis dynamics is  dominated by the longitudinal component (also on the level of the gradient).
 \subsection{Volkov-type states in arbitrarily structured laser field}
\label{sec:SFA}
Let us consider the simplest example of a unbound electron subject to a strong structured laser field.
In the case of spatially homogeneous vector potential such a state is described by a Volkov wave \cite{wolkow1935klasse}.
Analytical (Volkov-like) solutions for the unbounded electron motion in generally structured fields were not reported sofar. Below, we derive under certain conditions "structured-light Volkov wave"  (SL-VW).
From the discussion so far  and considering Fig.\,\ref{fig1}c-d, we conclude that reasonable approximations   should capture the action of the vector potential gradient $\pmb{\nabla}\pmb{A}^{m^{(a)},\sigma_{\rm L}}_{\rm OV}$.\\
An atom at the position $\pmb{r}_0=(x_0,y_0,0)^T$ in the laser focal plane  experiences
a vector potential at $\pmb{r}_0$ that varies  smoothly in space. Suppressing for clarity sub and superscripts of $\pmb{A}$  and Taylor expanding around $\pmb{r}_0$ yields   for the $j$-th component to a first order
${A}_j(\pmb{r},t)={A}_j(\pmb{r}_0,t) + \sum_i {r}_i {M}_{ij}$ or equivalently (${r}_i$ is $i^{th}$ component of $\pmb{r}$)
\begin{equation}
\pmb{A}(\pmb{r},t)=\pmb{A}(\pmb{r}_0,t) + \pmb{r}\cdot\doubleunderline{M}(t).
\label{eq:Aloc}
\end{equation}
The matrix elements of  $\doubleunderline{M}(t)$ are $M_{ij}= \left. \partial_{r_i} {A}_j(\pmb{r},t) \right|_{\pmb{r}=\pmb{r}_0}$. The treatment of the first order term enables the inclusion of non-dipolar effects \cite{walser2000high, chirilua2002nondipole}.  In the presence of  $\pmb{A}(\pmb{r},t)$, the Hamiltonian of an electron  bound by the potential
$V(\pmb r)$   reads then
\begin{equation}
\hat{H}(t)=\frac{1}{2}\left[\hat{\pmb{p}} + \pmb{A}(\pmb{r}_0,t) + \pmb{r}\cdot\doubleunderline{M}(t) \right]^2 + V(r).
\label{eq:Ht}
\end{equation}
With $\pmb{\mathcal{E}}=-\partial_t\pmb{A}(\pmb{r}_0,t)$ being  the electric field and using the gauge transformation $|\Psi_L\rangle = e^{i\pmb{r}\cdot\pmb{A}(\pmb{r}_0,t)}|\Psi\rangle$ one obtains
\begin{equation}
\hat{H}_L(t)=\frac{1}{2}\left[\hat{\pmb{p}}+ \pmb{r}\cdot\doubleunderline{M}(t) \right]^2 + \pmb{r}\cdot\pmb{\mathcal{E}}(t) + V(r),
\end{equation}
 Noting that  $[\hat{\pmb{p}},\pmb{r}\cdot\doubleunderline{M}(t)]_-=0$,
and neglecting  higher order terms in the local variation of $\pmb{A}$ (i.e., $[\pmb{r}\cdot\doubleunderline{M}(t)]^2\approx 0$) we write
\begin{eqnarray}\label{eq:volkov13}
\hat{H}_L(t)&=& \hat{H}_{\rm Volkov}(t) + V(r), \nonumber\\
\hat{H}_{\rm Volkov}(t)&=&\frac{1}{2}\hat{\pmb{p}}^2 + \pmb{r}\cdot\doubleunderline{M}(t)\cdot\hat{\pmb{p}} + \pmb{r}\cdot\pmb{\mathcal{E}}(t).
\end{eqnarray}
There is an opportunity for a  nonperturbative analytical  treatment if considering  $\hat{H}_{\rm Volkov}(t)$ and $V$ to act separately, which is the basis of the strong field approximation  \cite{faisal1973multiple,keldysh1965ionization,reiss1980effect}  (strong means that the field terms dominates $V$ when considering the unbound electron dynamics). Such an approximation  is worthwhile doing, for a series of important  phenomena and experiments can be
described reasonably well within this strong field approximation \cite{Amini_2019}.   For us the key issue here is to find the function $|\Psi_{\pmb{p}}^{\rm (V)}(t)\rangle$ obeying
\begin{equation}\label{eq:volkovH}
i\partial_t|\Psi_{\pmb{p}}^{\rm (V)}(t)\rangle = \hat{H}_{\rm Volkov}|\Psi_{\pmb{p}}^{\rm (V)}(t)\rangle.
\end{equation}
To derive the expression for this  state which we termed above SL-VW,  one may proceed at first as for the conventional Volkov state by writing the ansatz  \cite{wolkow1935klasse, pisanty2018high}
\begin{equation}
|\Psi_{\pmb{p}}^{\rm (V)}(t)\rangle = e^{-\frac{i}{2}\int^t\pi^2(\pmb{p},\tau){\rm d}\tau}|\pmb{\pi}(\pmb{p},t)\rangle.
\label{eq:PsiVolkov}
\end{equation}
The states $|\pmb{\pi}(\pmb{p},t)\rangle$ are plane waves propagating with the kinematic momenta $\pmb{\pi}(\pmb{p},t)$, meaning $\hat{\pmb{p}}|\pmb{\pi}(\pmb{p},t)\rangle= \pmb{\pi}(\pmb{p},t)|\pmb{\pi}(\pmb{p},t)\rangle$.
Thus, Eq. (\ref{eq:volkovH}) amounts to integrating
\begin{equation}
\frac{\partial\pmb{\pi}(\pmb{p},t)}{\partial t} + \pmb{\mathcal{E}}(t) = - \doubleunderline{M}(t)\cdot\pmb{\pi}(\pmb{p},t).
\end{equation}
Recalling that  $\int_t\doubleunderline{M}(t){\rm d}t\sim(q/\omega)\pmb{A}(\pmb{r}_0,t)=(1/c)\pmb{A}(\pmb{r}_0,t)$,
we write $\pmb{\pi}(\pmb{p},t)=\pmb{p}+\pmb{A}(\pmb{r}_0,t) + \delta\pmb{\pi}(\pmb{p},t)$ and seek a solution to first order in (1/c) which yields
\begin{equation}
\pmb{\pi}(\pmb{p},t)=\pmb{p}+\pmb{A}(\pmb{r}_0,t) - \int^t{\rm d}\tau \doubleunderline{M}(\tau)\cdot(\pmb{p}+\pmb{A}(\pmb{r}_0,\tau)).
\label{eq:kin_mom}
\end{equation}
  As detailed below, even in regions where $\pmb{A}(\pmb{r}_0,t)$  is very small  $\doubleunderline{M}$ may be large enough such that the last term in eq.(\ref{eq:kin_mom}) may even dominate the behaviour of the Volkov phases ($A_j$ and gradient of ${A}_j$ are independent). Such a case is encountered when an atom resides in the vicinity of the optical vortex  core in the parallel class, described by $\pmb{A}_{\rm OV}^{m^+,+1}$ or $\pmb{A}_{\rm OV}^{m^-,-1}$.\\
  The key quantity of SL-VW is its
 phase   $S_V(\pmb{p},t,\pmb{r}_0)=\frac{1}{2}\int^t {\rm d}\tau\,\pmb{\pi}(\pmb{p},\tau,t')^2$, or explicitly \\
\begin{widetext}
\begin{equation}
S_V(\pmb{p},t,\pmb{r}_0)
      =E_pt + \frac{1}{2}\int^t{\rm d}\tau\pmb{A}^2(\pmb{r}_0,\tau) +\pmb{p}\cdot\int^t{\rm d}\tau\left[ \pmb{A}(\pmb{r}_0,\tau)-\int^\tau{\rm d}t'' \doubleunderline{M}(t'')\cdot(\pmb{p}+\pmb{A}(\pmb{r}_0,t''))\right]
\label{eq:SV}
\end{equation}
\end{widetext}
where $E_p=p^2/2$. The second term is related to  the action of the local ponderomotive potential
(terms  containing higher powers of $\pmb{A}(\pmb{r}_0,t)$ are neglected). In principle, having Eq.(\ref{eq:SV}) one may in retrospect  insert the determined SL-VW into Eq.(\ref{eq:volkovH}), and assures the consistency of the approximations. The explicit form of the SL-VW depends on the type of the vector potentials and is discussed below for some typical cases.
  \subsection{Electrons in a strong  OAM carrying  laser field}
%
Let us consider the phase  of SL-VW  for the case where an optical vortex of the parallel class  acts  on an electron  that has been released  from an atom  residing at $\pmb{r}_0=(\rho_0\cos\varphi_0,\rho_0\sin\varphi_0,0)^T$. It reads
\begin{widetext}
\begin{equation}
S_V^{(m^+,+1)}(\pmb{p},t,\pmb{r}_0)= \frac{1}{2}\left(p^2 +  A_0^2(q_\perp\rho_0)^{2m}\right)t + \alpha_m\sin\vartheta_{\pmb{p}}\left[\frac{m}{\rho_0} \frac{pq_z}{\omega q_\perp}\sin\vartheta_{\pmb{p}}\cos(2\varphi_{\pmb{p}}-\omega t) - \left(1 + \frac{q_zp\cos\vartheta_{\pmb{p}}}{\omega}\right) \sin(\varphi_{\pmb{p}}-\omega t) \right]
\label{eq:OV1}
\end{equation}
\end{widetext}
We expressed $\pmb{p}$ with its amplitude $p$ and the spherical angles $\vartheta_{\pmb{p}},\varphi_{\pmb{p}}$. In Eq.(\ref{eq:OV1}) $\alpha_m=A_0p(q_\perp\rho_0)^m/\omega$ characterizes the displacement of the electron at the position $\pmb{r}_0$ in the  structured laser field.
 Note that $\doubleunderline{M}(t)\cdot\pmb{A}(\pmb{r}_0,t) \propto qA_0^2 $ and was therefore  neglected.
 In the antiparallel case which occurs  for instance for a topological charge $m^-$ and $\sigma_{\rm L}=+1$ (opposite chiralities of SAM and OAM), the most relevant contributions to the SL-VW phase are
\begin{widetext}
\begin{equation}
\begin{split}
S_V^{(m^-,+1)}(\pmb{p},t,\pmb{r}_0)=&\frac{1}{2}\left(p^2 + A_0^2(q_\perp\rho_0)^{2m}\right)t + \alpha_m\left[ (-1)^{m+1} \frac{m}{\rho_0} \left(\frac{2}{q_z}\cos\vartheta_{\pmb{p}} +
\frac{p}{2\omega}(1+3\cos^2\vartheta_{\pmb{p}})\right)\cos\omega t \right. \\
&\left. - \left(1 + \frac{q_zp\cos\vartheta_{\pmb{p}}}{\omega}\right)\sin\vartheta_{\pmb{p}} \sin(\varphi_{\pmb{p}}-\omega t) \right].
\end{split}
\label{eq:OV2}
\end{equation}
\end{widetext}
The influence of the orbital angular momentum $m$ of the laser fields enters the SL-VW for both cases as terms which scale as $m/\rho_0$. For very large (compared with $q_z^{-1}$) axial distances $\rho_0=\sqrt{x_0^2+y_0^2}$, differences between Eq.\,\eqref{eq:OV1} and Eq.\,\eqref{eq:OV2} vanish, and the SL-VW converges to the conventional Volkov wave for spatially uniform circularly polarized light. This is to be expected, as $m$ is related to the optical axis. Hence, for an atom at large $\rho_0$, the phase of the vector potential is basically constant. This observation can be exploited for spatially resolved photoemission on the scale below the optical diffraction limit: Photoelectrons that show dependence on $m$ must have started from regions around the optical axis, or in general from regions where the spatial phase of the vector potential varies significantly on the scale of the atomic wave functions \cite{watzel2016discerning}. This argument may also serve for using the photoelectrons to map the structure of the optical fields, as demonstrated below.
  The independence of the SL-VW given by (\ref{eq:OV1},\ref{eq:OV2})  on the atom-position polar angle $\varphi_0=\arctan(-y_0/x_0)$ reflects the symmetry of the considered system (the vector potential and the atom).
  \subsection{Electrons driven by polarization textured vector beams}
For an unbound electron in a  vector beam with azimuthal polarization we find for  the SL-VW phase  the form
\begin{widetext}
\begin{equation}
\begin{split}
S_V^{\rm AVB}(\pmb{p},t,\pmb{r}_0)=&\frac{1}{2}\left(p^2t + (A_0q_\perp\rho_0)^2 \right)t + \alpha_1\left( 1 + \frac{pq_z}{\omega}\cos\vartheta_{\pmb{p}}\right) \sin\vartheta_{\pmb{p}}\sin(\varphi_{\pmb{p}}-\varphi_0)\cos(\omega t).
\end{split}
\label{eq:AVB}
\end{equation}
\end{widetext}
For a radially polarized vector beam the expression is markedly different  encompassing the influence of the   longitudinal component which becomes more important for tighter focusing.  We infer the expression
\begin{widetext}
\begin{equation}
\begin{split}
S_V^{\rm RVB}(\pmb{p},t,\pmb{r}_0)=\frac{1}{2}\left[p^2 + A_0^2\left(q_\perp^2\rho_0^2 + \frac{q_\perp^2}{q_z^2}\left(4-q_\perp^2\rho_0^2\right)\right)\right]t
+ \alpha_1&\left[ \vphantom{\frac{2}{q_z\rho_0}} \left(1-\frac{q_\perp^2\rho_0^2}{8}+\frac{pq_z}{\omega}\cos\vartheta_{\pmb{p}}\right) \sin\vartheta_{\pmb{p}}\cos(\varphi_{\pmb{p}}-\varphi_0)\sin(\omega t) \right. \\
&\left.
- \frac{2}{q_z\rho_0}\left(1-(q_\perp\rho_0)^2/4\right) \cos\vartheta_{\pmb{p}}\cos(\omega t)\right].
\end{split}
\label{eq:RVB}
\end{equation}
\end{widetext}
\subsection{Electron quantum dynamics  in intense  optical skyrmionic fields}
In the case of an optical propagating  skyrmion, some properties of the  modified Volkov phase can be inferred from the two  vortices with different topological charges $m_1^+$ and $m_2^-$ that form the skyrmion but the cylindrical symmetry cannot be exploited as in previous cases and the expression is thus more involved:
\begin{widetext}
\begin{equation}
\begin{split}
S_V^{\rm OS}(\pmb{p},t,\pmb{r}_0)=&\frac{1}{2}\left[p^2 + \left(\alpha^2(q_\perp\rho_0)^{2m_1} + \beta^2(q_\perp\rho_0)^{2m_2} \right)\, A_0^2\right]t\\
& + \frac{A_0p\sin\vartheta_{\pmb{p}}}{\omega_{\rm L}}\left[ \vphantom{\frac{A_0p\sin{\vartheta}}{\omega_{\rm L}}} \beta(q_\perp\rho_0)^{m_2}
\sin[m_2(\pi+\varphi_0)+\varphi_{\pmb{p}} + \omega_{\rm L}t] - \alpha(q_\perp\rho_0)^{m_1}
\sin[m_1\varphi_0 + \varphi_{\pmb{p}} - \omega_{\rm L}t] \right] \\
& + \frac{A_0p^2\sin^2\vartheta_{\pmb{p}}}{\omega_{\rm L}^2}\left[ \vphantom{\frac{A_0p\sin{\vartheta}}{\omega_{\rm L}}} \alpha \frac{m_1}{\rho_0}(q_\perp\rho_0)^{m_1}
\cos[2\varphi_{\pmb{p}} - \omega_{\rm L}t - (m_1-1)\varphi_0] \right. \\
&\quad\quad\quad\quad\quad\quad\quad
\left.+ \beta \frac{m_2}{\rho_0}(q_\perp\rho_0)^{m_2}
\cos[2\varphi_{\pmb{p}} + \omega_{\rm L}t - m_2(\pi+\varphi_0) -\varphi_0] \right].
\end{split}
\label{eq:OSphase}
\end{equation}
\end{widetext}
The meaning of the various terms entering  $S_V^{\rm OS}(\pmb{p},t,\pmb{r}_0)$ follows from the discussions of Eq.(\ref{eq:OV2}), as the skyrmionic field receives contributions from two vortices with winding numbers $m_{1,2}$. Whether terms associated with $m_1$ or $m_2$ are locally more important  depends on the ratio
$(\alpha(q_\perp\rho_0)^{m_1})/(\beta(q_\perp\rho_0)^{m_2})$.
\section{Applications}
Having derived the electronic wave function in the presence of a structured  intense laser field, we utilize it for the description of selected physical processes, namely a) photoionisation assisted by structured intense laser fields, b) for steering and momentum texturing of electronic wave packets with radially polarized vector beam, and c) for spatio-temporal mapping of skyrmionics optical fields. For concreteness we use in all calculations below a He atom as a typical target. The potential $V$ confining the electrons to the atom is modelled within the effective single-particle approach, discussed and mathematically detailed in Ref. \cite{tong2005empirical}.
\subsection{Photoionization of atoms assisted by intense optical vortices}
\begin{figure}[t!]
\centering
\includegraphics[width=0.97\columnwidth]{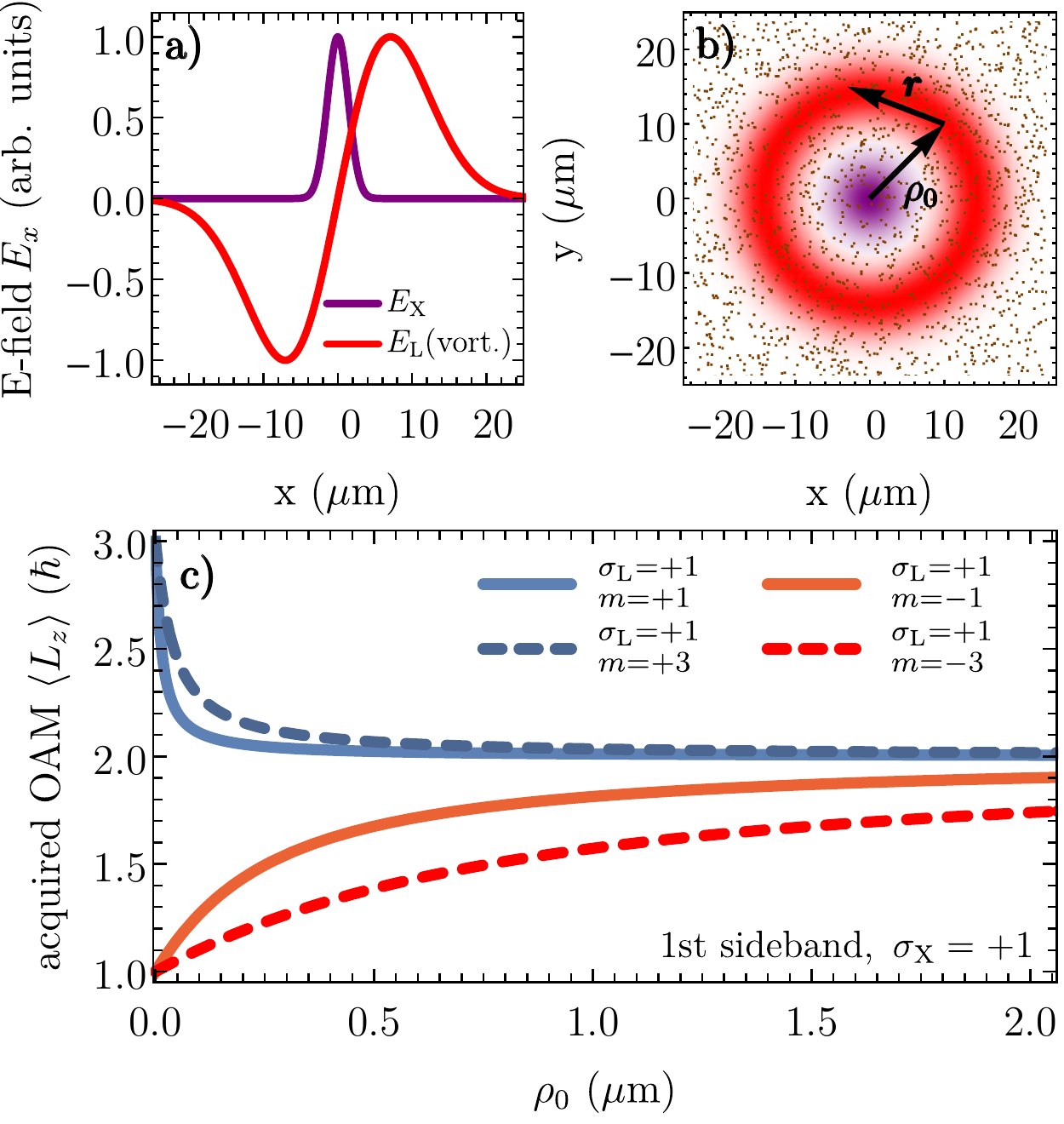}
\caption{XUV photoionization  assisted with an optical vortices laser pulse. a) Spatial distributions of both laser fields in the focal plane where (b) a gas of atoms is located. The extent of the interaction region is set by the profile of the X-ray field, which ionizes an atom directly upon absorption of one photon. c) Acquired OAM by the photoelectron for circularly polarized optical vortices with winding numbers $m_{\rm OAM}=\pm1$ (solid lines) and $m_{\rm OAM}=\pm3$ (dashed lines). The detected energy belongs to the first sideband.}
\label{fig:fields}
\end{figure}

Let us consider the liberation of a valence shell electron upon the  absorption of one (X)UV photon with energy $\hbar\omega_{\rm X}$. In addition, an intense   structured laser field $L$ with frequency $\hbar\omega_{\rm L}$ is present. This laser $L$ affects strongly the photoelectron wavepacket dynamics in a way that can be quantified as follows:
The photoionization amplitude in the presence of the two laser fields reads \cite{faisal1973multiple, milovsevic2019atom}
\begin{equation}
\mathcal{A}_{\pmb{p}}(\pmb{r}_0)=-i\int_{-\infty}^{\infty}{\rm d}t'\langle\Psi^{-}_{\pmb{p},\rm SV}(\pmb{r}_0,t')|\hat{H}_{\rm X}(t')|\Psi_i(t')\rangle.
\label{eq:amplitude0}
\end{equation}
Here, $\hat{H}_{\rm X}(t')$ is the interaction Hamiltonian of the valence shell electron with the (X)UV field and $|\Psi_i(t)\rangle$ is the initial state. The time-dependent final state is
\begin{equation}
|\Psi^{(-)}_{\pmb{p},\rm SV}(\pmb{r}_0,t)\rangle=e^{i[\pmb{A}(\pmb{r}_0,t)-\pmb{K}(\pmb{p},t)]\cdot\pmb{r}} |\Psi^{(-)}_{\pmb{p}}\rangle e^{-iS_V(\pmb{p},t,\pmb{r}_0)}.
\label{eq:Psi_ASV}
\end{equation}
where $\pmb{K}(\pmb{p},t)=\int^t{\rm d}\tau\,\doubleunderline{M}(\tau)\cdot(\pmb{p}+\pmb{A}(\pmb{r}_0,\tau)$ and $|\Psi^{(-)}_{\pmb{p}}\rangle$ satisfies  the time-independent Schr\"{o}dinger equation for the atomic Hamiltonian $\hat{H}_{\rm at}=\hat{\pmb{p}}^2/2 + V(r)$ for the kinetic energies $E_p>0$. The Volkov phase $S_V(\pmb{p},t,\pmb{r}_0)$ is given in Eq.\,\eqref{eq:SV}. In contrast to Eq.\,\eqref{eq:PsiVolkov}, we use the full scattering states in the atomic potential $V(r)$ instead of plane waves.
The justification of using the final states $|\Psi^{(-)}_{\pmb{p},\rm SV}(t)\rangle$ follows the formal steps when deriving  the Coulomb-Volkov ansatz, given in Refs.\,[\onlinecite{faisal2016strong,milovsevic2019atom}].
Similarly, it can be shown that Eq.\,\eqref{eq:amplitude0} is the zeroth order amplitude corresponding to the integral equation describing the time evolution operator
\begin{equation}
\hat{U}(t,t')=\hat{U}_{\rm SV}(t,t') - i\int_{t'}^t{\rm d}\tau\,\hat{U}_{\rm SV}(t,\tau)\hat{V}_{\rm SV}(\tau)\hat{U}(\tau,t')
\label{eq:TimeEvolution}
\end{equation}
which involves  the full Hamiltonian $\hat{H}(t)$ [cf.\,Eq.\,\eqref{eq:Ht}]. We note that $|\Psi^{(-)}_{\pmb{p},\rm SV}(t\rightarrow\infty)\rangle=|\Psi^{(-)}_{\pmb{p}}\rangle$, where $t\rightarrow\infty$  is the time  when measurement is conducted (at the photoelectron detector) while all  laser fields are off.\\
%
%
While the formulation applies to all types of structured fields, we select the case when
the assisting laser field L is an infrared intense (IR) optical vortex carrying OAM and it propagates collinearly with the homogeneous (on the scale of the atoms), weaker XUV field, a situation which has been experimentally realized recently \cite{de2020photoelectric, mazza2016angular}. Our focus here is on the theoretical aspects. Details of the experiments and comparison with theory are discussed at length in Ref.\cite{de2020photoelectric} where the target was a thermal gas cell of helium atoms \cite{meyer2010two}.\\
During detection, via the photoelectron energy selection we may zoom to those events where one photon from the XUV laser and one IR photon are involved. Interestingly, we may even achieve a time ordering on which photon is absorbed first by choosing a tightly focused XUV and less focused IR laser beam, in which case our photoelectrons first absorb the XUV photon in the region where the IR laser has a very low (or vanishing) intensity and then experience the IR laser on their way out to the detector (cf. Fig.\,\ref{fig:fields}a-b ). This scenario implies also a spatial resolution on the position of the involved atom on the scale of the laser spot of the XUV laser. In a way, our setting  resembles the case of STED-microscopy (STED=stimulated emission depletion) \cite{Westphal246}. In fact, if we would investigate few XUV photon processes (that we can select via the photoelectron energy), we would increase the spatial dependence to around the intensity peak of the XUV laser.\\
Generally, if we are interested in effects related to the spatial structure of the laser,
the photoelectron should be slow enough such that the first kinetic energy term in Eq.(\ref{eq:volkov13}) does not completely overwhelm the field terms.
Interestingly, the slow photoelectrons that take notice of the local phase structure of the IR laser, absorb the IR photon in the vicinity of the optical axis and not at the IR field maximum (where the local field resembles for the atom a Gaussian field).
Hence, our approach in deriving  Eqs.\,\eqref{eq:parallel1}-\eqref{eq:antiparallel2} is indeed useful.\\
For concreteness, we choose the XUV photon energy to be $\hbar\omega_{\rm X}=30$\,eV. The durations of both laser fields are assumed very long  (short pulses can be associated with the streaking regime \cite{kazansky2010angle}), in which case on both sides of the main-photonline, additional lines are well-developed and are separated by $\hbar\omega_{\rm L}=1.55$\,eV.\\
The spatial phase of the IR pulse is reflected in the  difference of the photoelectron yields corresponding to the  IR pulses with $m^+$ or $m^-$, meaning an OAM-induced dichroism. The conventional  circular dichroism and how it relates to target's orientation and/or alignment is well-established in the literature (for example, in Refs.\cite{kazansky2011circular, mazza2014determining} and references therein), similar arguments apply to circularly polarized optical vortices \cite{baghdasaryan2019dichroism, seipt2016two}. The circular dichroism in photoexcitation by using optical vortices was presented in Ref.\,\cite{afanasev2017circular}, where transitions involving higher multipolarity revealed a strong difference. \\
We consider an XUV field with a fixed helicity of $\sigma_{\rm X}=+1$ so that the ejected photoelectron is orbitally oriented.
The XUV laser field spatial distribution is assumed to be Gaussian $f(\rho_0)=e^{-(\rho_0/(2w_{\rm X}))^2}$, where $w_{\rm X}$ is the effective width of the focal spot.
In the rotating wave approximation \cite{landau2013quantum}, the X-ray interaction Hamiltonian can be expressed as $\hat{H}_{\rm X}(t)=H_{\rm X}e^{-i\omega_{\rm X}t}$ with $H_{\rm X}\propto rY_{1,1}(\Omega_{\pmb{r}})$, which can be inserted  into Eq.\,\eqref{eq:amplitude0}. \\
To trace the transfer of the optical OAM to the electrons, we investigate the angular momentum acquired by the photoelectron
\begin{equation}
\langle L_z\rangle = \frac{\langle\psi_{\rm SB}|\hat{L}_z|\psi_{\rm SB}\rangle}{\langle\psi_{\rm SB}|\psi_{\rm SB}\rangle},
\end{equation}
where $\hat{L}_z=-i\partial_\varphi$ and the wave function belonging to a specific side band (SB) is found by the projection
\begin{equation}
\psi_{\rm SB}(\pmb{r},t)=\int_{\rm SB}{\rm d}\pmb{p}\,\mathcal{A}_{\pmb{p}} \Psi_{\pmb{p}}^{(-)}(\pmb{r})e^{-iE_pt}.
\end{equation}
Here, the integration is performed (numerically) around the $n$th sideband's energy $E_n=\hbar(\omega_{\rm X} + n\omega_{\rm L})+E_i$, i.e. $E_p\in[E_n-\epsilon,E_n+\epsilon]$ where $\epsilon$ is determined by the energy width of sideband. \\
 Fig.\,\ref{fig:fields}c) shows the acquired OAM of the photoelectrons for final energies in the first sideband (i.e., one IR vortex photon is absorbed) depending on the IR laser field's winding number and on                                              the atom's distance $\rho_0$ to the optical axis. Let us inspect the case $m=1$: At $\rho_0\rightarrow0$, the OAM transfer converges against $(m^{(a)}+\sigma_{\rm L}+\sigma_{\rm X})\hbar$, meaning that  in the parallel case ($m^+,\sigma_{\rm L}=+1$) the  vortex field boosts the angular momentum of the photoelectron. This can be explained by the modified selection rules \cite{picon2010photoionization}, i.e., $|1s^2\rangle\xrightarrow{\rm X}|Y_{11}\rangle\xrightarrow{\rm L}|Y_{33}\rangle$ by absorption of one photon from each the X and the L laser fields. For the antiparallel case ($m^-,\sigma_{\rm L}=+1$), $|1s^2\rangle\xrightarrow{\rm X}|Y_{11}\rangle\xrightarrow{\rm L}\alpha|Y_{31}\rangle+\beta|Y_{11}\rangle$ indicating that no total angular momentum is transferred to the photoelectron.\\
Increasing the axial distance $\rho_0$, we verify that  $\langle L_z\rangle$ converges to $(\sigma_{\rm L}+\sigma_{\rm X})\hbar=2\hbar$, i.e. the L laser field s locally homogeneous and circularly polarized (vortex' helicity $\sigma_{\rm L}$ is spatially independent).\\
 We note  the different "decay rates" of the OAM transfer, which can be traced back  to the spatial components of the vector potential: In the \emph{antiparallel} case, the decay is slower due to the emergence of the longitudinal component. However, OAM transfer in the \emph{parallel} case
is mediated by the first derivative of $\pmb{A}_{\rm OV}^{m^+,\sigma_{\rm L}=+1}$, whose effective distance is limited. In Fig.\,\ref{fig:fields}c), we demonstrate that the decay rates can be slowed down by increasing the topological charge as presented by the dashed curves. We recall however that the behavior of the curves for $m>1$ and $\rho_0\rightarrow0$ is not  correctly described by our theory in the full range, because  the theory accounts up to the first derivative in Eq.\,\eqref{eq:Aloc}. For $m>1$, higher derivatives of order $m$ are necessary for a correct description (the transverse component scales like $\rho^m$). In those cases however, the ionization probabilities at the origin are negligibly small, which justifies our restriction to the first order in the series expansion of $\pmb{A}_{\rm OV}$.
\subsubsection{OAM-dependent dichrosim}
The  theory presented so far proves the OAM transfer in vortex laser-assisted photoionization. On the other hand, in a photoionisation experiment typically differential cross sections (DCS) $\propto\mathcal{W}(\pmb{p})$ are measured.
\begin{figure}[t!]
\centering
\includegraphics[width=0.97\columnwidth]{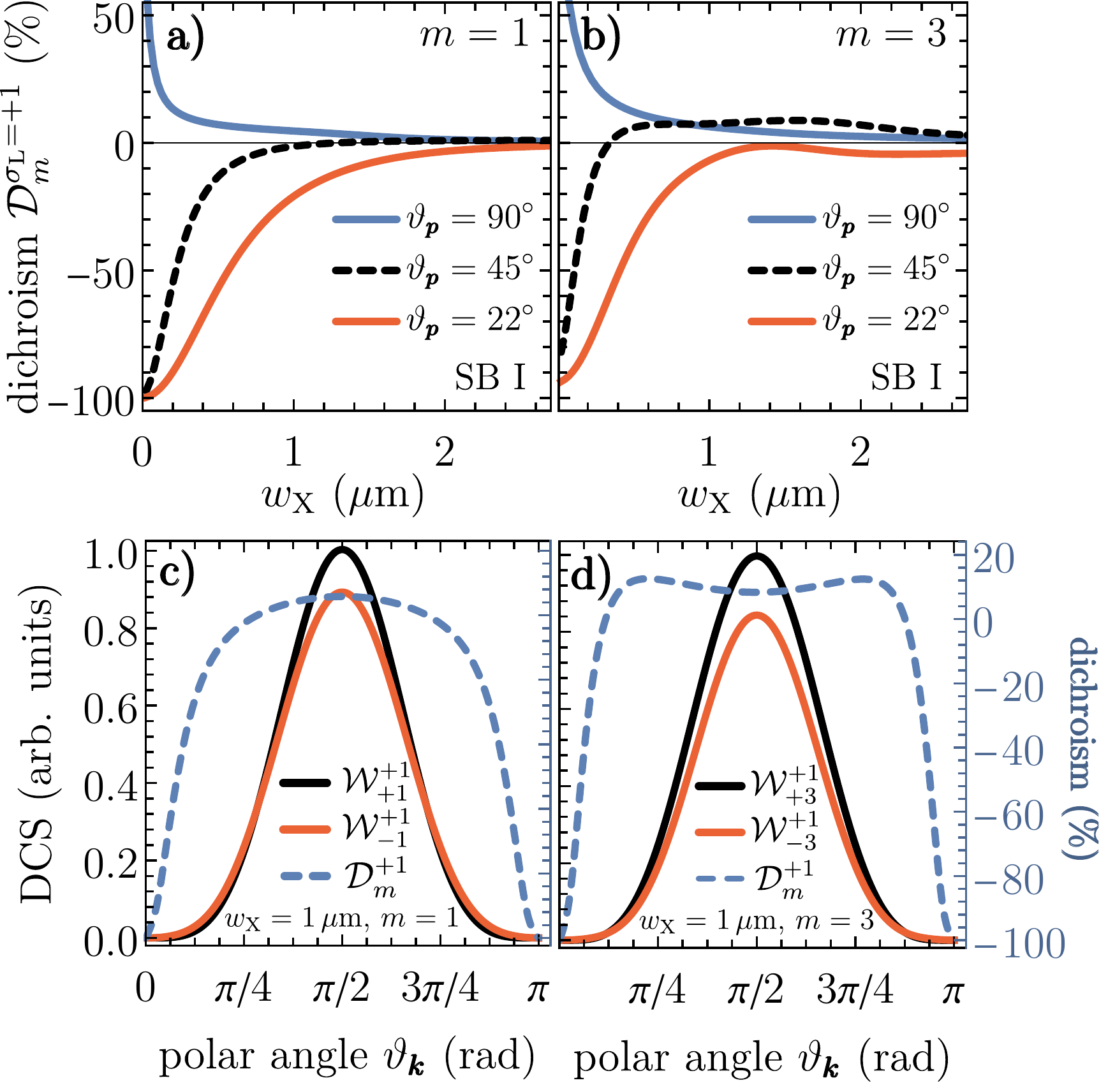}
\caption{Dichroism in the photoionization probability corresponding to circularly polarized optical vortices. a) Dichroism  dependence on the width $w_{\rm X}$ of the ionizing X-ray field spot for $m=1$. b) the same as a) for $m=3$. c) Angular-resolved, averaged (over X-ray spot) DCS and dichroism for $w=1\,\mu$m and $m=1$. d) the same as c) for $m=3$. }
\label{fig3}
\end{figure}
The question is then whether the laser-matter OAM transfer may show up in  angular or energy-resolved  ionization probability. To quantify the answer, two different measurements with fixed SAM state $\sigma_{\rm L}$ and pulse parameters ($A_0, w_{\rm L}$) are mandatory. On this basis, we define the orbital dichroism $\mathcal{D}^{\sigma_{\rm L}}_{m}$ as the normalized difference in the (measured) photoelectron yields using two OAM-carrying lasers that have oppositely directed orbital chiralities, while the polarization state is fixed. More precisely, we define:
\begin{equation}
\mathcal{D}^{\sigma_{\rm L}}_{m}=\frac{ \mathcal{W}_{m^+}^{\sigma_{\rm L}}(\pmb{p}) - \mathcal{W}_{m^-}^{\sigma_{\rm L}}(\pmb{p}) }{ \mathcal{W}_{m^+}^{\sigma_{\rm L}}(\pmb{p}) + \mathcal{W}_{m^-}^{\sigma_{\rm L}}(\pmb{p}) }.
\end{equation}
Here, the ionization probabilities are statistically averaged over a macroscopic distribution of atoms (gas sample)
\begin{equation}
\mathcal{W}^{\sigma_{\rm L}}_{m}(\pmb{p})=2\pi\int_0^\infty{\rm d}\rho_0\,\rho_0\left|\mathcal{A}_{\pmb{p}}^{m,\sigma_{\rm L}}(\rho_0)\right|^2.
\end{equation}
In Fig.\,\ref{fig3}, we present the results for the dichroism in the laser-assisted photoionization by circularly polarized optical vortices ($\sigma_{\rm L}=+1$). We concentrate here on the first sideband in the electron spectra, which corresponds to the absorption of one  photon from the vortex laser L yielding a continuum-continuum transition.
Panel a) shows the dichroism depending on the size of the interaction region (determined by the width $w_{\rm X}$ of the ionizing laser (X) field for different (asymptotic) directions $\vartheta_{\pmb{p}}$ of the photoelectrons in the case of $m=1$. The limited (spatial) range of the dichroism is ubiquitous: similar to the OAM transfer mediated by the modified Volkov phases, we observe a fast decay of the dichroism in all directions by increasing the effective interaction region. The dichroism represents the different actions of the transversal and longitudinal field components: the ionization probability $\mathcal{W}_{m^+}^{\sigma_{\rm L}=+1}(\pmb{p})$ belongs to the vortex in the parallel class (cf.\,Eq.\,\eqref{eq:parallel1}) while
$\mathcal{W}_{m^-}^{\sigma_{\rm L}=+1}(\pmb{p})$ is associated with the antiparallel class, where the longitudinal component dominates the dynamics near the optical axis. Therefore, it is not surprising that we find a \emph{positive} dichroism in the transversal plane, which we can trace back to the transversal field component of the parallel class vector potential $\pmb{A}_{\rm OV}^{m^+,\sigma_{\rm L}=+1}$. Detecting, however, the photoelectron more in the direction of the light propagation axis, i.e., in the vicinity of $\vartheta_{\pmb{p}}=0$, results in a \emph{negative} dichroism, which we attribute to the action of the longitudinal component present in the antiparallel vector potential $\pmb{A}_{\rm OV}^{m^-,\sigma_{\rm L}=+1}$. Hence, similar trends as in the OAM transfer can be inferred: the angular dependence of the dichroism can be related to a smooth transition between the short-ranged effect of the transversal component (around the transverse plane) and the long-ranged effect of the longitudinal component. \\
Fig.\,\ref{fig3}c) shows the angular resolved DCS and the corresponding dichroism for an interaction region $w=1\,\mu$m. Both probabilities peak in the transverse plane, which is usual for laser-assisted photoionization.The dichroism is positive around $\vartheta_{\pmb{p}}=\pi/2$  changing  sign rapidly when the photoelectrons emerge near the optical axis.\\
Figures.\,\ref{fig3}b-d) present the same results for a higher winding number, i.e., $m=3$. Increasing the vortex' carried orbital momentum increases the dichroism and the range, which is particularly apparent when comparing the blue curves, belonging to $\vartheta_{\pmb{p}}=\pi/2$, between Figs\,\ref{fig3}a) and \ref{fig3}b). Furthermore, the domain where $\mathcal{D}_{m=3}^{\sigma_{\rm L}=+1}>0$ is  increased. This is in line with our observation of the acquired angular momentum, as highlighted in Fig.\,\ref{fig:fields}a) and by the inspection of Eqs.\,\eqref{eq:OV1}-\eqref{eq:OV2}. The terms representing the impact of the OAM are proportional to $m/\rho_0$ so that increasing the topological charge enhances the effect.\\
The results so far  underline that the OAM transfer can be linked to the different behaviors of the corresponding photoionization probabilities, particularly when compared to the photoelectron yield in the transverse plane.
Increasing the photoelectron's angular momentum by the absorption of a vortex photon with a suitable OAM direction (i.e., parallel to the photon's helicity) results in an enlarged photoionization probability in the transverse plane. The absorption of a vortex photon carrying \emph{antiparallel} OAM (relative the helicity) decreases the cross-section in transverse direction giving rise to a dichroism.
\subsubsection{Coherence and thermal average}
\label{sub:coh}
As schematically depicted in Fig.\ref{fig:fields}b), the atoms are stochastically distributed in the laser spot (red ring in Fig.\ref{fig:fields}b) and have an extension way below the optical wavelength. The spatial phase of the OAM-carrying laser is related to the spatial angular coherence of the laser wavefront. Any spatial fluctuations of the laser phase blurs the value of the carried OAM. So how comes that a phase of a classical field defined on such a length scale can be imprinted on thermally distributed, extremely localized electronic quantum state \cite{de2020photoelectric}.   In principle, we may pose the same question regarding the sensitivity of photoelectrons to circular polarization of a homogeneous field, for such a polarization is nothing but the coherent oscillation (in time) of two independent (but equal in strength)  transverse components which are phase shifted by $\pm\pi/2$.  This phase shift is everywhere the same. Each of the independent and thermally distributed atoms reacts hence locally in the same way to this phase shift and therefore the thermal average does not affect the circular dichroism. This same argument applies to the spatial phase of the laser. What is constant here (for OAM carrying fields for instance) is the angular gradient of the vector potential (which is proportional to $m$). The additional caveat however is that the radial distribution is not homogeneous (in contrast to the case of a circular polarization) and $m$ is defined with respect to the optical axis. On the other hand,   right on the optical axis the intensity is very low or vanishing so that the light-matter interaction is very weak. Substantially away from the optical axis the atoms are insensitive to $m$ \cite{PhysRevA.86.063812}. Thus, contributions to the orbital dichroism in $m$ stem from a narrow ring around the optical axis (whose radius for weak fields is discussed in  \cite{PhysRevA.86.063812}) where the OAM-transfer is  independent of the atom's angular position and hence unaffected by thermal averaging. In fact, a denser gas cell is more favorable for an experimental observation (note the absorption-emission time is instantaneous on the scale of the thermal atomic motion).
\begin{figure}[t!]
\centering
\includegraphics[width=0.95\columnwidth]{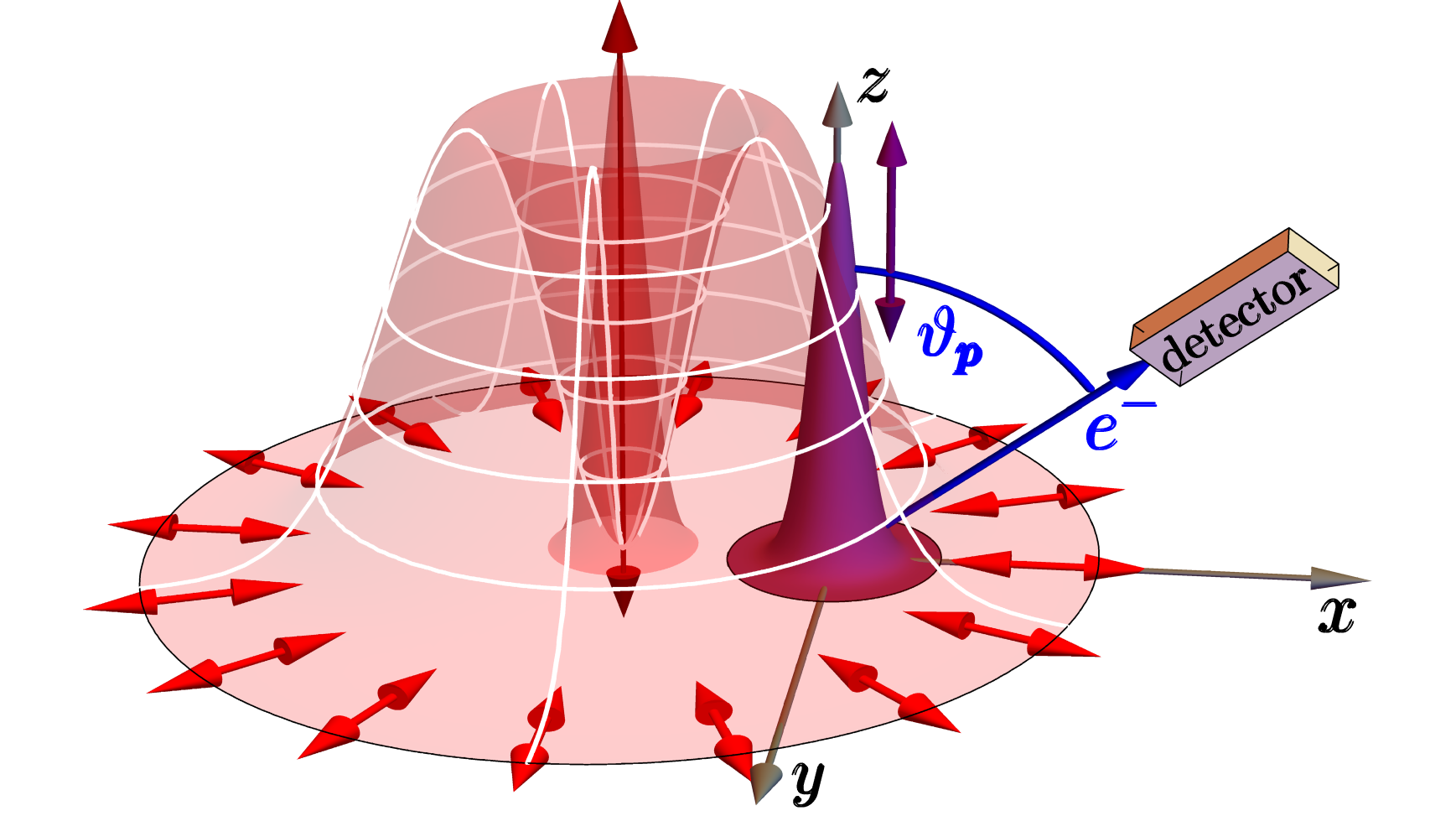}
\caption{Schematic representation of laser-assisted photoionization setup with a tightly focussed RVB. A focused $z$-linearly polarized XUV field ionizes a valence shell electron into the external IR RVB.  Depending  on the region from which the electronb wave packet is launched, the photoelectron experiences the distribution of the longitudinal and the transversal field components of the vector beam. The double-headed arrows represent the local polarization directions of the RVB fields.}
\label{Fig_repr_RVB}
\end{figure}
\subsection{Steering and momentum-texturing of electronic   wave packet  via radially polarized  vector beams}

\begin{figure}[t!]
\centering
\includegraphics[width=0.95\columnwidth]{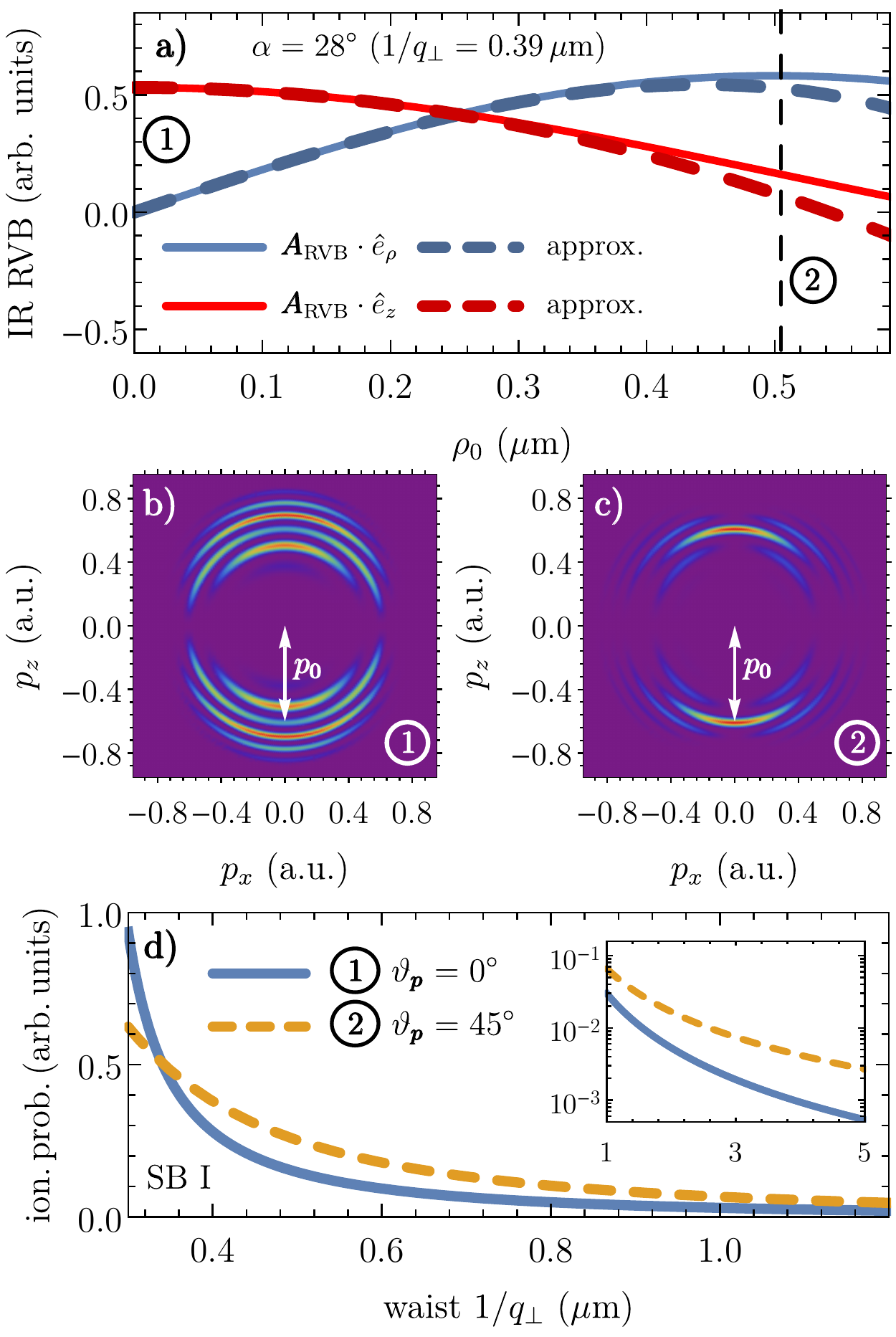}
\caption{Laser-modified photoelectron wavepacket motion in  a tightly focussed RVB. a) The dependence of the longitudinal and transversal field components   on the axial distance $\rho_0$. b-c) Photoionization spectra in the dependence on the final momenta $p_x$ and $p_z$ for two different launching positions within the laser spot. $p_0$ belongs to the sole absorptions of an XUV photon with $\hbar\omega_{\rm X}=30$\,eV (zeroth side band). Higher side bands belong to absorption/emission of several IR photons. d) Photoionization probability for a fixed asymptotic direction within the first side band (SB I) as a function of  the extent of the laser spot (given by the waist $w_{\rm L}=1/q_\perp$).}
\label{figRVB}
\end{figure}
By focusing a radially polarized vector beam we can realize a transition from a dominating longitudinal component in the vicinity of the optical axis to a transversal component for outer radii \cite{watzel2020multipolar}. This transition occurs on the sub-wavelength scale. An example of the individual field components is shown in Fig.\,\ref{figRVB}a).\\
Depending on the atom's position within the laser spot, the liberated photoelectron is exposed to an external laser field with different spatial components and varying (local) amplitude, as presented schematically by Fig.\,\ref{Fig_repr_RVB}. For a demonstration, we picked up two positions, which are represented by 1 and 2 in Fig.\,\ref{figRVB}. By absorbing IR photons of the assisting vector beam, the photoelectron wave packet is accelerated, which is visible in the (measured) momentum distribution. Moreover, the direction of the acceleration is crucially dependent on which field component is locally dominating.\\
In the following example, the ionizing XUV field is $z$-linearly polarized with $\hbar\omega_{\rm X}=30$\,eV, while the focussing of the assisting RVB is determined by $\tan\alpha=q_\perp/q_z$ ($\alpha=28^\circ$). For this condition, the IR laser spot size's diameter in the focal plane is around the wavelength $\lambda_{\rm L}$. A rather high amplitude of $A_0=0.15$\,a.u. together with an infinitely long pulse lengths enable the emergence of higher side bands \cite{kazansky2010angle}, as highlighted by Figs.\,\ref{figRVB}b-c. A photoelectron liberated from an atom located in the vicinity of the optical axis is exposed to the strong longitudinal component of the external IR vector beam, and - upon absorption of IR photons, it is accelerated into the propagation direction. Moreover, boosting the photoelectron wave packet's kinetic energy results in a pronounced concentration of ionization probability around $\vartheta_{\pmb{p}}=0$. As mentioned before, the relative strengths between the longitudinal and the transversal field components of the tightly focused RVB vary very quickly on a sub-wavelength scale. Hence, for an atom located further away from the optical axis, the measured photoelectron's characteristics changes strongly. If the electron wave packet is ejected into the $z$-direction by the XUV field, the  transversal component forces the photoelectron into a transverse  trajectory, as illustrated in Fig.\,\ref{figRVB}c, where higher side bands corresponding to the exchange of several IR photons are visible. The spectrum reveals that the  center of the respective angular-dependent ionization probability wanders more and more towards the vicinity of $\vartheta_{\pmb{p}}=\pi/2$. \\
Therefore, the direction of the photoelectron wave packet's acceleration can be manipulated by the vector beam components via  focusing: Keeping all the laser parameters unchanged, the photoionization probability exhibits  a strong dependence on the laser spot size determined by $w_{\rm L}=1/q_\perp$. This connection is demonstrated in Fig.\,\ref{figRVB}d), where the ionization probability for the first side band (SB I) is presented for two fixed asymptotic directions and positions of the atoms. Interestingly, for a tightly focused RVB, $\mathcal{W}(p_1,\vartheta_{\pmb{p}}=0)$ (belonging to 1) related to the longitudinal component is larger than the one (position 2) related to the  transverse component. Broadening the beam waist changes the situation: While both probabilities decrease with increasing $w_{\rm L}$, the individual decay rates are different. As a consequence and as a general rule, for larger extents of the focal spots, interaction with the assisting vector beam is much more probable via the transversal field component. For weakly focused RVBs, with  a waist  in the range of 10 microns, the acceleration due to the longitudinal component is practically not present.
\subsubsection{Coherence and role of atom spatial distribution}
In Sec.\,\ref{sub:coh}, we argued why the OAM transfer to a stochastic atom distribution is not washed away by configurational averaging. For AVB and RVB the situation has some subtleties. For a strongly focused RVB, the longitudinal component is dominant and its action resembles the case of a linear polarized field (along the propagation direction). Therefore, the response of the atom distribution in the focus is linearly related to the response of a single atom (we suppressed so far the well-established propagation and phase-matching issues related to a finite length (along $z$ direction) of the sample).
For de-focused vector beams, the intensity distribution is similar to vortex beams. Let us for concreteness consider AVB, as given by Eq.~\ref{eq:avb}. Alternatively, we can also write for this beam $\pmb{A}_{\rm AVB}(\pmb{r},t) = i/(2\sigma_{\rm L})\; A_0q_\perp\rho( \hat{e}_{-\sigma_{\rm L}} e^{i\sigma_{\rm L}\varphi} - \hat{e}_{\sigma_{\rm L}} e^{-i\sigma_{\rm L}\varphi})\sin(q_zz-\omega_{\rm L}t)$. The coherence in the time oscillations of the transverse field components are reflected by the value of $\hat{e}_{\sigma_{\rm L}}$. In AVB  (or RVB) $\hat{e}_{\sigma_{\rm L}}$ is everywhere the same except for a fixed spatial  rotation angle as $\varphi$ evolves (signified by $e^{\pm i\sigma_{\rm L}\varphi}$ in  $\pmb{A}_{\rm AVB}(\pmb{r},t) $). Therefore, the response of a strongly inhomogeneous atomic distribution will be different from that of a statistically distributed one. Indeed, this fact is reflected in the non-trivial dependence of the  SL-VW phases in Eqs.\,\eqref{eq:AVB} and \eqref{eq:RVB}  on the atom angular position $\varphi_0$ from which the electron is launched. Clearly, one can reverse  the argument and retrieves from the photoelectron distributions information on the spatial structure of the atom distribution in the vector beam laser spot, on a scale below the optical wavelength. Similar arguments also apply to optical skyrmions.
\begin{figure}[t!]
\centering
\includegraphics[width=0.9\columnwidth]{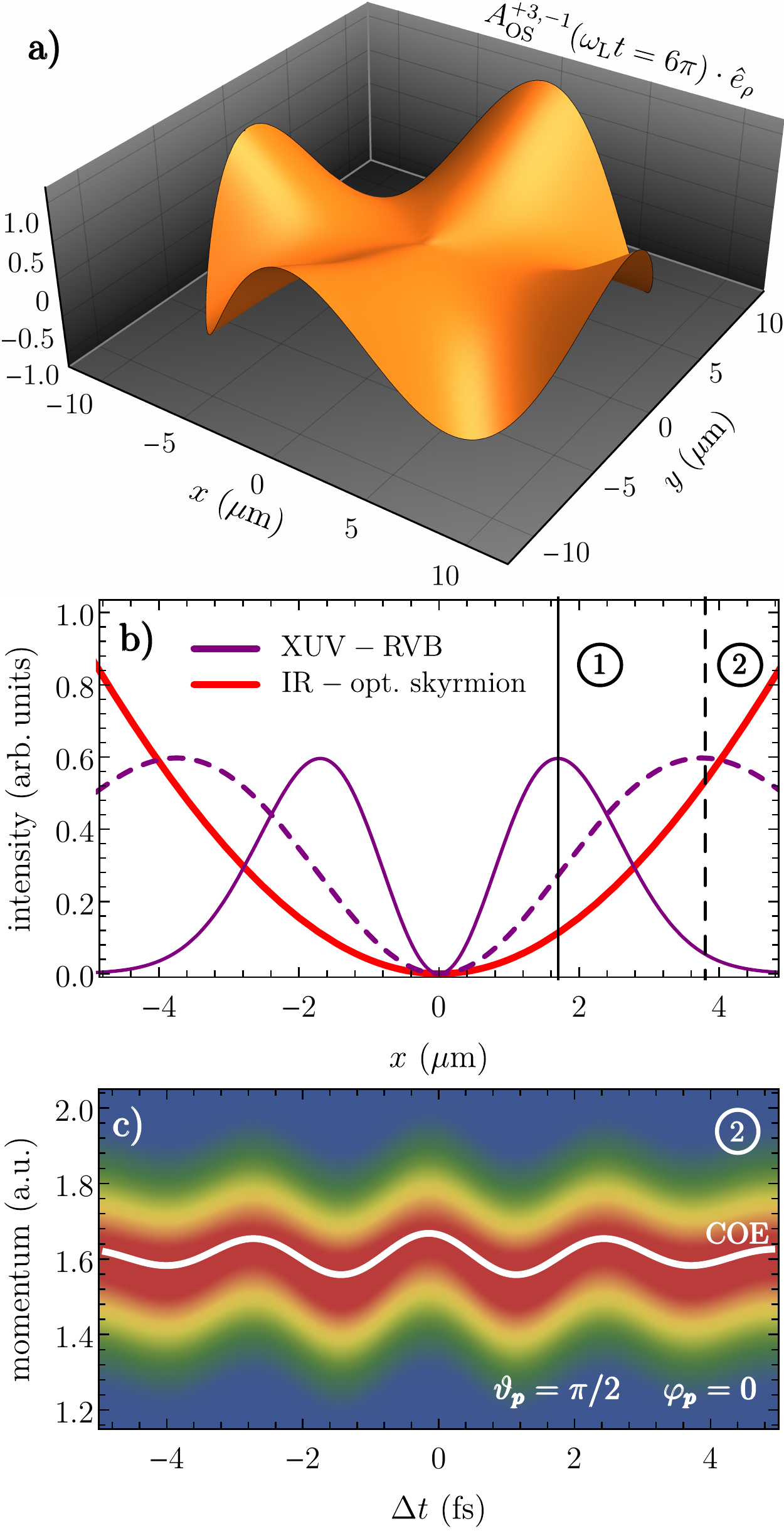}
\caption{Photoelectrons traversing an optical skyrmion. a) Radial component of the vector potential in the focal plane. b) Setup: An XUV radial vector beam with variable focusing liberates an photoelectron at its intensity maximum with high probability, where the local spatio-temporal IR optical skyrmion is present. (1) and (2) present two different XUV focusing. c) Streaking spectrum for a photoelectron detected in the asymptotic direction $\vartheta_{\pmb{p}}=\pi/2$ and $\varphi_{\pmb{p}}=0$. The center of energy (COE) represents the maximum of the photoionization probability  depending on the delay time $\Delta t$.}
\label{fig4}
\end{figure}
\subsection{Reconstruction of an optical propagating skyrmionic field via attosecond streaking}
Let us consider as an example the optical skyrmion $\pmb{A}_{\rm OS}^{m_1=3,m_2=-1}(\pmb{r},t)$ ($\alpha=7$, $\beta=1$) with a moderate  focusing of $w_L=7\,\mu m$. The corresponding radial component of the vector potential is shown in Fig.\,\ref{fig4}a), revealing a strong azimuthal variation. Note that the optical skyrmion is shown only in the  range where our approximation of the spatial distribution function $F_m(\rho)$ is valid. Our goal is to sample the local structure of the skyrmionic field via a traversing electronic wave packet.  Generally, one may use the attosecond streaking technique \cite{goulielmakis2004direct, krausz2009attosecond, pazourek2015attosecond} which has been established as a key element of attosecond spectroscopy. If the XUV laser field is a short pulse, its large bandwidth allows for several quantum paths, involving absorption and emission of several IR photons, to a final energy state.
\begin{figure}[b!]
\centering
\includegraphics[width=0.95\columnwidth]{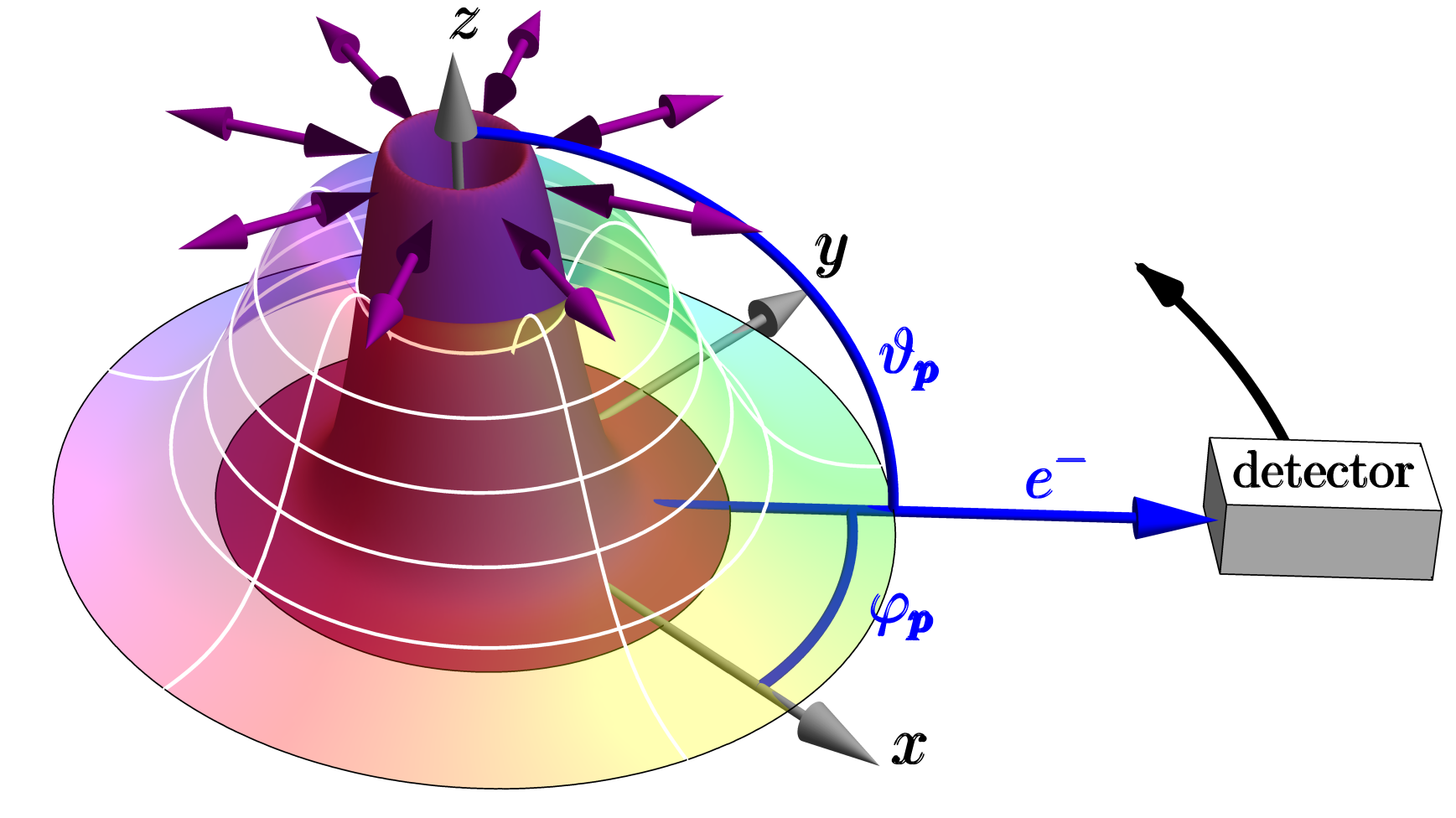}
\caption{Schematic representation of the light field - detector geometry. Photoelectrons are primarily emitted in the focal plane due to the radially polarized XUV field and are streaked by the local IR optical vortex which is characterized by a varying polarization and phase landscape. The Photoelectrons are measured in directions $\{\vartheta_{\pmb{p}}=\pi/2,\varphi_{\pmb{p}}\}$.}
\label{fig4a}
\end{figure}
Interference between those quantum paths results in a modulation of the final momentum, which depends on the temporal delay $\Delta t$ between both pulses (in our case, $\Delta t$ refers to the temporal difference between the maxima of both pulses) \cite{dahlstrom2013theory}. Classically, the detected momentum of the photoelectron (in a specific asymptotic direction $\Omega_{\pmb{p}}=\left\{\vartheta_{\pmb{p}},\varphi_{\pmb{p}}\right\}$) follows $p(\Delta t)\approx\sqrt{2(\omega_{\rm X}+E_i)}-\tilde{A}_{\rm L}(\Delta t)$, where $\tilde{A}_{\rm L}$ is the projection of the vector potential in the direction of the (measured) asymptotic momentum. For low amplitudes of the external laser (L) field, the above-mentioned classical relationship is an excellent approximation.\\
The temporal relation between the photoelectron's final momentum and the vector field's amplitude at the moment of ionization allows for the imaging of the optical skyrmion using the photoemission streaking measurements.
If the photoemission spectra are recorded in every asymptotic direction $\varphi_{\pmb{p}}$ in the focal plane ($\vartheta_{\pmb{p}}=\pi/2$), a phase shift should be visible since the streaking field (the optical skyrmion) has an internal phase structure along the azimuthal direction. In other words, for fixed delay times $\Delta t$, the final asymptotic momentum in the presence of both laser fields becomes explicitly directionally dependent, reflecting the unique phase structure of the optical skyrmion, as shown in Fig.\,\ref{fig4}a). Moreover, the phase variation of the structured field depends on the axial distance.
\begin{figure}[t!]
\centering
\includegraphics[width=0.95\columnwidth]{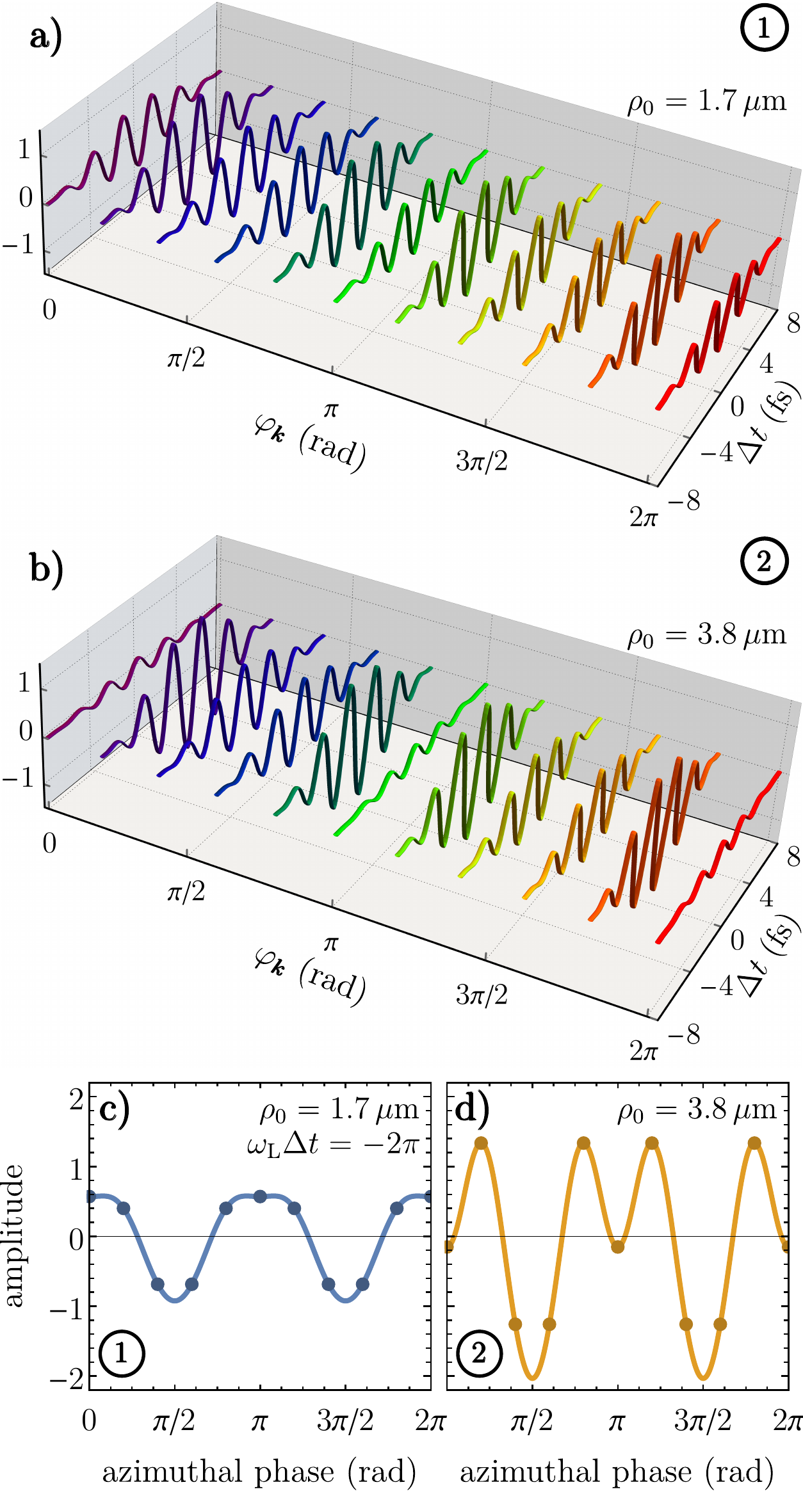}
\caption{Reconstructed information retrieved from the photoemission spectra. a-b) Directional dependent COEs for two different focusing setups of the ionizing XUV RVB revealing the temporal information about the local field of the IR optical skyrmion. c-d) Reconstructed phase variation for a fixed delay time $\Delta t$. The curves follow from an interpolation of the data points with periodic boundary conditions.}
\label{fig5}
\end{figure}
To deal with this feature, one may use a radially polarized vector beam \cite{hernandez2017extreme}, as shown schematically in Fig.\,\ref{fig4}b). By adjusting the focus of the donut-shaped intensity distribution, we can select atoms  within the laser spot that are photoionized in the radial direction in the transverse plane (with respect to the optical axis). Only atoms near the intensity maximum of the XUV field have a sizable ionization probability via one-photon processes. Once liberated, the photoelectrons are affected (streaked) by the local field of the IR optical skyrmion. Measuring now the photoelectron in the asymptotic direction $\left\{\vartheta_{\pmb{p}}=\pi/2, \varphi_{\pmb{p}}\right\}$ exploits the radial field component $A^{+3,-1}_{\rm OS}\cdot\hat{e}_\rho$ of the optical skyrmion in the photoemission spectrum. A schematic representation is given in Fig.\,\ref{fig4a}.\\
Mathematically, we gain access to the streaking spectrum by numerical integration of Eq.\,\eqref{eq:amplitude0} and by calculating the modified Volkov phases for $A^{+3,-1}_{\rm OS}(\pmb{r},t)$, given in Eq.\,\eqref{eq:OSphase}. In addition, we introduce the delay time $\Delta t$ in the Hamiltonian describing the interaction between the XUV-photon and an atom located at the axial distance $\rho_0$ within the laser spot:
\begin{equation}
\hat{H}_{\rm X}(\rho_0,t-\Delta t)=\rho\mathcal{E}_{\rm RVB}(\rho_0)f(t-\Delta t)\cos[\omega_{\rm X}(t-\Delta t)].
\end{equation}
Here, $\mathcal{E}_{\rm RVB}(\rho_0)=E_0(\rho_0/w_{\rm X})\exp(-\rho_0^2/w_{\rm X}^2)$ is the radial distribution function of the vector beam, where $w_{\rm X}$ determines the focusing. The action of the longitudinal component can be neglected since the XUV beam is weakly focused [cf.\,Fig.\,\ref{fig4}b)]. The temporal envelope of the short pulse is given by $f(t)=\cos(\omega_{\rm X}t/(2n))^2$ for $t\in[-n\pi/\omega_{\rm X},n\pi/\omega_{\rm X}]$ (and zero otherwise). In our simulations, we chose $\hbar\omega_{\rm X}=60$\,eV and a number of $n=7$ optical cycles, which means we are in the streaking regime \cite{kazansky2010angle}. For the setups shown in Fig.\,\ref{fig4}b), $w_{\rm X}=2.4\,\mu$m (case 1) and $w_{\rm X}=5.3\,\mu$m (case 2) were chosen. The fixed peak intensity (at the maximal field amplitude) of $2\times10^{14}$\,W/cm$^2$ ensures that the single photon processes are initiated by the radially polarized XUV field.\\
In Figs.\,\ref{fig5}a-b), we present the center of energy (COE), extracted from the corresponding streaking spectra, for different asymptotic directions $\varphi_{\pmb{p}}$ and two different focusing setups of the ionizing XUV pulse. Due to the spatial extent of the XUV field, we can "scan" the IR field characteristics in the radial direction, whereas the azimuthal dependencies can be retrieved from rotating the photoelectron detector [cf.\,Fig.\,\ref{fig4a}]. By varying $\varphi_{\pmb{p}}$, the COEs reveals modulation both in the amplitude and on the time axis. This is, in particular, evident in the second focusing setup presented in Fig.\,\ref{fig5}b). A further important  aspect is the $2\pi$-periodicity revealing the imprinting of the optical field's phase onto the photoelectron distributions. By fixing the time delay $\Delta t$, we can reconstruct the phase information of the optical skyrmion, as shown in Figs.\,\ref{fig5}c-d). As anticipated, the phase structure of a skyrmion is very involved due to the interplay of the two contributing vortices with different orbital  angular momenta. By varying  the focusing  of the XUV field (i.e., adjusting the waist parameter $w_{\rm X}$), the full radial field component of the optical skyrmion can be scanned. The other spatial components $A^{+3,-1}_{\rm OS}\cdot\hat{e}_\varphi$ and $A^{+3,-1}_{\rm OS}\cdot\hat{e}_z$ can retrieved from Maxwell's equations. Usually, plasmonic optical skyrmions are mapped via PEEM (photoemission electron microscopy) \cite{tsesses2018optical,Spektor1187,dai2019ultrafast,PhysRevX.9.021031}. The complementary method proposed here can also map the spatio-temporal structure of freely propagating optical  skyrmions.

\section{conclusions}
The goal has been to derive  quasi-analytical expressions for the electron motion in an intense laser field that exhibits a non-trivial distribution in space of the spin and/or of the wavefront. Besides, we aim at exploiting the derived expression for describing physical processes such as laser-assisted ionization of atoms, particle acceleration, or spatio-temporal mapping of topological photonic fields. The well-known Volkov wave fully describes the motion of an unbound electron in an unstructured laser field. In contrast, the structured laser field Volkov state (SL-VW) we were able to obtain only under certain approximations that can be in principle improved systematically, but such improvements were not treated here. Fortunately, for a number of physical processes of interest, the derived approximate SL-VW is valid and useful. Generally, the SL-VW is fundamentally different from the conventional Volkov state since it receives contributions from the spin-angular momentum spatial distribution as well as from the space structure of the vector potential. Notably, also the field scalar potential affects SL-VW, and its influence can be encapsulated in a longitudinal component of the vector potential.  As a demonstration, we considered photoionization assisted by propagating optical vortices, meaning a field with a well-defined spatial phase structure (but no spin-angular momentum structure). Orbital angular momentum of the laser field is related to this phase and is found to be transferable to the photoelectrons, even for atoms that are stochastically distributed in the laser spot. The assisting laser pulse may carry no orbital angular momentum but its spin angular momentum can be structured, as for radially polarized laser pulse. In this case  we found that this beam can be employed to manipulate the momentum distribution of the electronic wavepacket. Thereby, the longitudinal component of the laser's vector potential is decisive. A further application in photonics  concerns topology. An electronic wavepacket traversing a topologically no-trivial optical field such as optical skyrmions samples spatio-temporal information on the skyrmionic field which enables a mapping of the phase and the spin-angular momentum texturing in the skyrmion, even if it is not in the form of a localized plasmonic field. Rich important phenomena results from applying a combination of time delayed or frequency shifted structured fields. For example,  applying a RVB and a time delayed linearly polarized pulse to a torus generates a field-free toroidal moment as (excited) electronic eigenstates   \cite{Watzeltoroid}.   In addition to  application in electronics and magnetism, future applications include understanding high harmonic emission in arbitrarily structured fields as well as field-assisted particle trapping and stabilization.
\section{Acknowledgements}
This work was supported by the DFG through SFB TRR 227 and WA 4352/2-1.
%

%

\end{document}